# In search of the origin of long-time power-law decay in DNA solvation dynamics

SAUMYAK MUKHERJEE[†], SAYANTAN MONDAL[†], SUBHAJIT ACHARYA, BIMAN BAGCHI*

Solid State and Structural Chemistry Unit
Indian Institute of Science, Bangalore – 560012, India

**GRAPHICAL TOC**

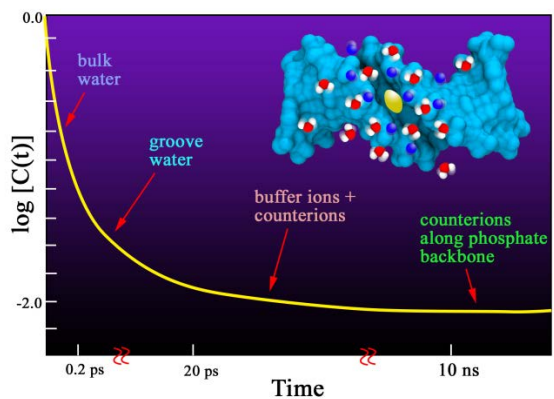

**Corresponding author**

*B Bagchi (Email - bbagchi@iisc.ac.in)

**Author contribution**

[†] S. Mukherjee and S. Mondal have contributed equally to this paper.

# ABSTRACT

Experiments reveal that DNA solvation dynamics (SD) is characterized by multiple time scales ranging from a few ps to hundreds of ns and in some cases even up to μs. The last part of decay is slow and is characterized by a power law (PL). The microscopic origin of this PL is not yet clearly understood. Here we present a theoretical study based on time dependent statistical mechanics and computer simulations. Our investigations show that the primary candidates responsible for this exotic nature of SD are the counterions and ions from the buffer solution. We employ the model developed by Oosawa for polyelectrolyte solution that includes effects of counterion fluctuations to construct a frequency dependent dielectric function. We use it in the continuum model of Bagchi, Fleming and Oxtoby only to find that it fails to explain the slow PL decay of DNA solvation dynamics. We then extend the model by employing the continuous time random walk technique developed by Scher-Montroll-Lax. *This approach can explain the long time PL decay, in terms of the collective response of the counter ions*. From MD simulations we find frequent occurrence of random walk of tagged counter ions along the phosphate backbone. We propose a generalized random walk model for counterion hopping and carry out kinetic Monte Carlo simulations to show that the nonexponential contribution to solvation dynamics can indeed arise from dynamics of such ions. We also employ a Mode Coupling Theory analysis to understand the slow relaxation that originates from ions in solution. *Explicit evaluation suggests that buffer ion contribution could explain logarithmic time dependence in the ns time scale, but not a power law*. From MD simulations we find log-normal distributions of relaxation times of water dynamics inside the grooves. This is responsible for the initial faster multiexponential decay of SD.

## I. INTRODUCTION

Dynamics of biomolecules are essential for their functions, such as enzyme kinetics, DNA transcription, drug-DNA intercalation etc. However, our understanding of biomolecular dynamics is still in its infancy, although much progress has been achieved in recent years.[1-6] Dipolar solvation dynamics has been developed into a useful probe to study both collective and local dynamics.[7-9] Initial applications were made to neat liquids, with several remarkable results, the most notable being the ultrafast solvation dynamics uncovered by Fleming and co-workers in the 1990s.[10-12] The origin of the ultrafast solvation has been attributed to the collective orientational relaxation that is driven by the large force constant of solvent polarization free energy surface, and aided by ultrafast modes like libration and rotation.[12-13] The relevant free energy surface was already envisaged in the celebrated Marcus theory of electron transfer reaction.[14] From the inception, understanding results of solvation dynamics experiments needed both reliable theory and computation.

In recent years, the technique of solvation dynamics has been used to study many complex systems, like organized assemblies, proteins and DNA.[15-19] In all these cases, solvation dynamics experiments have given rise to interesting results which have triggered much discussion leading to substantially better understanding and new insight. Protein solvation dynamics has been discussed extensively in recent times.[20-23] Solvation dynamics of DNA in particular has given rise to anomalous results that have proven hard to explain.[4, 24] The anomalous results of DNA solvation dynamics is the subject of present study. Our focus shall be on the slow long-time decay, as the initial fast part, which is controlled largely by water, has been addressed in several earlier studies.

*DNA is a remarkably flexible molecule*, to quote Lehninger from his well-known text book of Biochemistry.[25] A DNA not only undergoes bending, twisting, base pair opening, rolling motions, but also undergoes local melting. The latter could be fairly frequent because a double stranded DNA (ds-DNA) can have a melting point of just about 40-50°C above the normal temperature.[26] DNA conformational fluctuations are clearly accompanied by displacements of water molecules and counterions whose response could be adiabatic in the sense that these faster species can follow the motions of the DNA groups. However, we are not aware of many studies that address the coupling of the conformational fluctuations of DNA with the surrounding water and counter ions. In the case of drug-DNA intercalation, such studies have been initiated.

The motions of counterions are of special interest here because these motions can directly couple to dipolar solvation dynamics. These counterions are partly present near the negatively charged phosphate ions and are quite mobile. Experiments suggest that about half (50%) of the counterions inhabit the phosphate groups whereas the other half remain dispersed in solution.[27-28] These are huge sources of polarization fluctuations whose role in DNA solvation dynamics seems to have been mostly ignored in recent discussions. The contribution of the counterions to the dielectric relaxation of a DNA aqueous solution was considered in an elegant treatment by Oosawa[29] who pointed out that the fluctuations in the occupancy of counterions around the negatively charged phosphate back bone can not only provide a large increment to the observed value of the dielectric constant of the medium but can also give rise to extraordinary slow dynamics.

Another player in the polarization fluctuations of the liquid is the ions that constitute the buffer. These ions can respond to the sudden creation or alteration of a charge distribution. This response again can be quite slow. A theory was recently developed that addresses this particular point.[30] As already mentioned, another candidate of slow dynamics comes from the possibility of local melting that provides a unique characteristic feature to DNA double helix. μs simulation of B-DNA has shown good agreement with experimental results.[31]

The study of DNA structural dynamics and surrounding water was pioneered by Berg and co-workers in a series of papers.[32-33] They studied several probes that were either intercalated into DNA (by removal of one base pair) or bound to the grooves. The power law decay was particularly strong in the case of the intercalated probe.[24] Zewail and co-workers studied solvation dynamics of *2-aminopurine*,[34] and also other probes bound to the grooves. The latter studies did not find a power-law decay which could be attributed to the ultrafast time resolution used and short duration of the study. This point has remained unclear.

One of the first simulation studies that addressed DNA solvation dynamics was initiated by Bagchi, Hynes and co-workers.[35] These studies focussed on the dynamics of water molecules. There was no signature of any slow power law decay. Water dynamics in the grooves slowed down appreciably, but remained exponential-like, with the longest time scales in the range of 100 ps or so, thus much shorter than the time scales observed in the power law regime. However, subsequent computer simulation studies by Sen and co-workers observed the cross-over from exponential-like to power law type decay behaviour.[4] The crossover occurs well past 100 ps which could be the reason for the failure to observe either in experiments of Zewail and in simulations of Bagchi and Hynes.

Recently a theoretical study proposed that the observed slow power law decay could arise from the ions that are present in DNA solutions as buffers are used to stabilize the DNA.[30] This slow relaxation arises from long ranged inter-ion correlations present in an electrolyte solution, and captured in Debye-Hückel theory of ion atmosphere. The long range ion-ion correlations give rise to a slow decay in the dynamic structure factor of ions which could couple to the solvation dynamics of a dipolar probe. However experimental studies failed to verify this mechanism because change in buffer did not affect the power law decay to any significant degree. However, it seems too early to rule out the contribution of the buffer ions to DNA solvation dynamics. It is likely that these ions contribute to solvation energy of DNA with a time scale of a few hundred ps – that is intermediate between the ultra-slow dynamics of tens of ns and ultrafast dynamics of sub-hundred fs.

In summary, we can count three major contributions to DNA solvation dynamics.

(i) Contribution of the surrounding water molecules. In the time scales of discussion here, this contribution occurs in the ultrafast time scale. There could be a slightly slow component because of the contribution from groove water molecules but nothing slower than a few tens of ps.

(ii) Contribution from ions present in the aqueous solution of DNA as buffers.

(iii) Contributions from counterions. As mentioned above, the contributions of the first two (water and buffer ions) have been considered previously, in theory and simulations. However, we are not aware of any contributions towards solvation dynamics from the counterions.

Solvation dynamics of a dipolar probe is unique is the sense that it is sensitive to the collective dynamics but at the same time contains an intermediate range length scale that makes it more sensitive to local dynamics.[9, 36] This scale dependence may introduce non-exponentiality in the solvation time correlation function.

The present theoretical investigation employs multiple approaches.

(i) First, we use the continuum model theory developed by Bagchi, Oxtoby and Fleming (BOF)[37] that provides an elegant expression of the solvation time correlation function in terms of the frequency dependent dielectric function. In order to obtain the later, we invoke the approach developed by Oosawa[29] to obtain the required frequency dependent dielectric function to be used in the continuum model. We used so computed dielectric function in the solvation time correlation expression given by BOF. This procedure, although predicted the short time dynamics correctly, fails to reproduce the power law decay due to an inherent limitation of continuum model based theories.

(ii) We then adopted a direct approach. We generalized the Oosawa model to include a continuous time random walk (CTRW) model of Scher, Montroll and Lax[38-39] that provides a power law description of a polarization current due to the counterions. The combined theory indeed shows a power law decay in a straight-forward fashion. However, the model assumed that the diffusion equation based approach of Oosawa to be replaced by a CTRW model.

(iii) We then proceeded to check the CTRW picture of diffusion of counter ions. Studies of tagged counter ions indeed revealed both the diffusion and the mechanism of waiting time distribution (WTD). We indeed observed the diffusive migration of counter ions between the phosphate backbones. We also observed trapping of the tagged ion between parallel backbones, and sojourn into bulk water and then return. These provide a clear picture of the waiting time distribution.

(iv) Having established the CTRW, we carry out a model simulation to describe the motion of counterions. This model simulation combines the ideas of Oosawa and Scherr *et al*. This model reproduces the highly non-exponential nature of the counter ion contribution to solvation dynamics.

(v) In order to understand the possible involvement of the buffer ions, we employ a previously developed mode coupling theory with a precise analysis of the time scales. The basic time scale of this theory is the ion atmosphere relaxation which ranges between 100 ps and a few ns for the concentration used in experiments.

(vi) Finally, we present results of atomistic molecular dynamics simulations that explore certain aspects of our theory, in particular the modified structure and dynamics of groove water molecules and also the time scale of the relaxation of groove water molecules.

We next proceed to provide a brief review of the experimental studies of this complex problem.

## II. OVERVIEW ON EXPERIMENTAL STUDIES

Because of their great importance in diverse biological processes where dynamics often play critical role, a large number of experimental and simulation studies of solvation dynamics of DNA have been carried out. These studies are usually carried out by using a small piece of DNA duplex (like dodecamer to hexadecamer) and selectively attaching a probe on to the DNA. These experimental studies seem to have produced somewhat contrasting viewpoints.[4] Here we briefly summarize the known results, and suggest that there is no contradiction among various results. We review results but with an eye on to the questions that we attempt to answer in the subsequent sections.

In DNA solvation, the commonly used experimental techniques are the transient absorption (TA), the fluorescence up-conversion (UPC) and time correlated single proton counting (TCSPC). These experiments monitor time dependent fluorescence Stokes shift (TDFSS) and mainly measure the solvation time correlation function $C(t)$.[8, 40]

$$C(t) = \frac{v_f(t) - v_f(\infty)}{v_f(0) - v_f(\infty)} = \frac{\Delta E(t)}{\Delta E(0)} \qquad [1]$$

Unfortunately, there are no intrinsic natural residues in DNA that can be used as a natural fluorescence probe (like tryptophan in proteins). Hence, an external probe is to be attached that could be either covalently linked (for example, *Coumarin-102*, *2-aminopurine* etc.) also known as *base-stacked*; or non-covalently linked (for example, Hoechst, DAPI etc.) also known as *groove-bound*.[4]

It is important to note that these different experimental methods often probe different time regimes of relaxation. Hence, the results could differ from one method to another. When we include all the studies, the following unified picture emerges. We find that the following three components dominate DNA solvation dynamics. (i) First, an omnipresent sub one ps ultrafast component dominates initial energy relaxation. Note that dipolar solvation in neat water also exhibit relaxation components at ~250 fs and 1-2 ps. (ii) Second, a somewhat ambiguous, less than 100 ps (say, 1-100 ps) component which could be non-exponential. (iii) Lastly, a substantially slower component that ranges from few hundred picoseconds to nanosecond or even microsecond timescale. The last part appears to be well-described by a temporal power law.

The sub 1 ps ultrafast component appears to be universal and carries approximately 50-60 % of the decay of total solvation time correlation function.[21-22] The relative amplitude of this ultrafast component depends on the system and probe and is sensitive to the method of detection but the time constant is more robust. This inertial component arises from the collective response of solvent around the probe, aided by the large dielectric constant of water, intermolecular vibrational mode of the hydrogen bond network, and librational modes. Whereas this component is unanimously found in simulations, experimental studies miss this component because of the limited time resolution of the laser pulses often employed.[23]

Several experiments have reported timescales that are less than ~100 ps. Zewail and co-workers used *2-aminopurine* as a probe to study solvation dynamics by employing UPC (from 100fs to 50ps). They retrieved two timescales of 1.5 ps and 11.6 ps.[41-42] They attributed the faster (1.5ps) timescale to the dynamics bulk water molecules and the slower (11.6 ps) timescale to the weakly bound slow water molecules in the DNA hydration layer. Zewail *et al.*, in another study, used Hoechst as a groove binding probe and monitored TDFSS for ~100ps. Their study again revealed a bimodal decay with timescales of ~1.4 ps and ~19 ps, quite similar to that of *2-aminopurine*.[43-44] This led to the conclusion that the position of the probe does not significantly change the nature of solvation time correlation function in the region less that 100 ps. Simulation studies by Bagchi *et. al.*, Corcelli *et al.,* and others find a near quantitative match with the experiments of Zewail.[17, 35] Dallman *et al.* used *2-hydroxy-7-nitrofluorene* as a probe and observed three timescales ~220 fs, ~2.3 ps and 18.7 ps using TA technique.[45] Sen *et al.* have also reported observation of picosecond timescales, by using DAPI and Hoechst as a minor groove binding probe.[4, 32, 46-47] However, as mentioned earlier, all of the experimental studies missed the ultrafast component because of the limited resolution of the laser pulse employed. Zewail et al. and simulations also missed the ultraslow component that we discuss below.

The existence of much slower components that ranges from 100ps to tens of nanosecond timescales has been observed. This ultraslow component is not present in protein hydration dynamics and has raised considerable curiosity. In a pioneering study Berg *et al.* reported for the first time the observation of such slow solvation dynamics of base stacked *Coumarin-102* by employing TCSPC, in the 100ps to 30ns domain. Their study revealed that the solvation time correlation function is bi-exponential in nature with two distinct time constants ~300 ps and ~13 ns.[33]

Later, the same group, with better and more sophisticated tools, proposed that the solvation dynamics shows a logarithmic relaxation instead of a bi-exponential decay.[48] Sen and co-workers observed similar nanosecond order time components (~1-16 ns) in case of minor groove bound DAPI and Hoechst for a variety of DNA structures with the effect of base pair mismatch.[4, 46-47]

The origin of such slow timescales is not clear at present. There are clearly several candidates and all of them may contribute together. The slow dynamics could appear as a result of long wavelength modes of conformational fluctuations in DNA and also the movement of counterions that can couple to solvation of a dipolar probe. Perez *et al.* showed from microsecond all atom MD simulation of Dickenson's-dodecamer B-DNA that partial and total base pair openings can give rise to time scales up to ~1ns. The same study revealed that the positively charged counterions might stay bound to minor groove for as long as ~10-15 ns.[49] These slow processes could contribute to the slow timescales.

Note that Berg *et al.* used all three techniques for *Coumarin*-102 from 40 fs to 40 ns in order to map out the total solvation dynamics of the probe.[24] They concluded that only a single power law ($\sim t^{-0.15}$) can describe to solvation time correlation function over a wide time range. As already mentioned, the origin of DNA slow solvation is unclear at present. As one cannot extract the different exponential timescales from the power law fit, we speculate the origin as a complex dynamics that arises from a coupled motion of water, counterions and DNA itself. The earlier simulation studies by Bagchi, Corcelli and others were unsuccessful in capturing the power law decay clearly because of the short simulation trajectory length (~15-20 ns).[17, 35] Latest experimental results by Sen *et al.* reveal that the power law dynamics is similar to that of Berg *et al.* for the initial 100 ps, but deviates after that in case of groove bound probe.[4] Sen and co-workers showed that for minor groove bound DAPI the dynamics adapts a faster exponential decay after initial 100 ps.

In this paper, we aim to understand the origin of the observed power law by employing stochastic theories of time-dependent statistical mechanics combined with analytical and simulation results.

## III. COMPONENTS OF AQUEOUS DNA SOLUTION

An aqueous solution of DNA consists of several components that may participate in the solvation dynamics. Double stranded DNA itself contains a negatively charged phosphate backbone along with intricately positioned hydrogen bonded stacked base pairs. The phosphate moieties are solvated by sodium counterions ($Na^+$). The structure of DNA is sensitive to pH of the solution.[50] Hence in experiments, pH of the solution needs to be maintained by adding buffer solutions like sodium acetate, sodium phosphate etc. These contribute a significant number of cations and anions to the system. However, the major portion of the system is occupied by the water molecules.

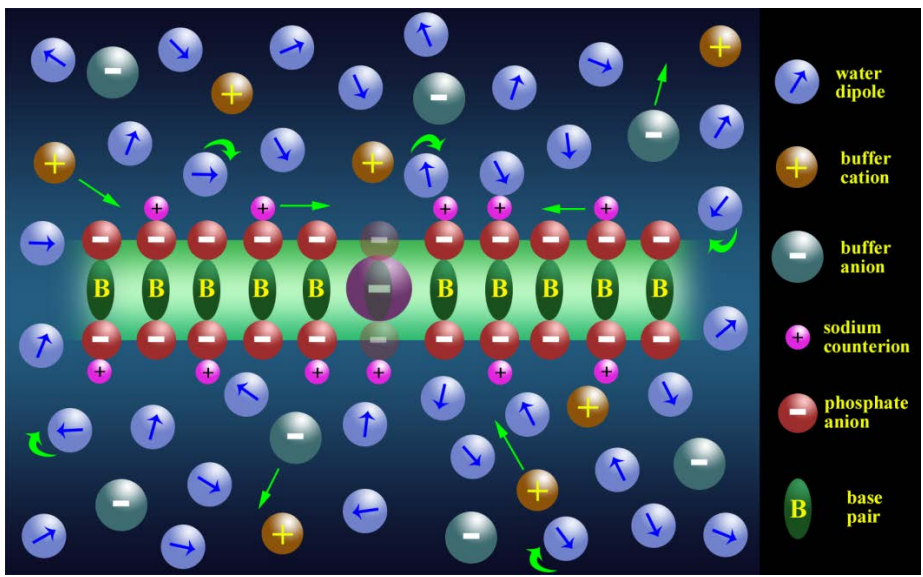

**Figure 1. Components of an aqueous DNA solution. Apart from the DNA (with base pairs and negatively charged phosphate backbones) and water, the system contains counterions and ions from the buffer solution. Solvation dynamics studies also involve a solute probe either intercalated to the DNA or bound to a groove. Green arrows represent a schematic translational or rotational movement of some of the components.**

In **Figure 1** we show a schematic representation of the crowded DNA solution described above, with an eye to the solvation problem being discussed here. Translational and rotational movements of some of the components are highlighted by green arrows. The solvation dynamics of such a system derives contributions from all the components. This introduces a spectrum of timescales in the dynamics of the whole system. Undoubtedly, the charged and dipolar species present in the system significantly affect the dielectric properties. We know that dielectric relaxation is intricately related to solvation dynamics. Hence a clear understanding of the individual contributions is required for a better interpretation of the dynamics of an aqueous DNA solution. In the following section (**Section IV**) we describe a theory that relates dielectric relaxation with solvation dynamics based on a continuum approach.

## IV. CONTINUUM THEORY OF SOLVATION DYNAMICS

In this section we use a continuum model approach to derive an analytical expression for the relationship between dielectric relaxation and solvation relaxation times. This derivation closely follows the work of Bagchi, Oxtoby and Fleming (BOF).[37] As already discussed, in experiments, solvation dynamics of a solute probe is measured by the time dependent fluorescence Stokes shift (TDFSS) of the emission spectrum.[51] The time-dependent energy decay [$\Delta E(t)$] of the probe is usually measured by the difference of average frequency at time $t$ [$\nu_f(t)$] from infinite time frequency $\left[\nu_f(\infty)\right]$. The normalized solvation time correlation function is defined by **Eq.[1]**

For our calculation we consider the solute probe as a cavity of radius $a$ with a point dipole (having a dielectric constant $\varepsilon_c$) at its center. Castner et al. has extended this treatment to include ellipsoidal cavity. The surrounding solvent environment is modeled by a dielectric continuum which is characterized by a frequency dependent dielectric constant $\varepsilon(\omega)$.[37] Hence the solute-solvent interaction is given by the interaction of the dipole inside the cavity with its own reaction field generated from polarization of the dielectric continuum. Under this approximation, the potential energy of the solute probe due to polar interactions is given by

$$\Delta E(t) = -\left(\frac{\varepsilon_c + 2}{3}\right)\boldsymbol{\mu}(t).\mathbf{R}(t) \qquad [2]$$

where $\boldsymbol{\mu}(t)$ is the dipole moment of the solute at time $t$ and $\mathbf{R}(t)$ is the time-dependent reaction field.[52] From a quasi-static boundary value calculation, the frequency-dependent reaction field $\mathbf{R}(\omega)$ is given by

$$\mathbf{R}(\omega) = r(\omega)\boldsymbol{\mu}(\omega) \qquad [3]$$

where,

$$r(\omega) = \left(\frac{2}{a^3}\right)\left[\frac{\varepsilon(\omega)-\varepsilon_c}{2\varepsilon(\omega)+\varepsilon_c}\right]\left(\frac{\varepsilon_c+2}{3\varepsilon_c}\right) \quad [4]$$

This equation is Fourier inverted to obtain the time-dependent reaction field (**Eq. [5]**).

$$\mathbf{R}(t) = \int_{-\infty}^{t} dt' r(t-t')\mathbf{\mu}(t') \quad [5]$$

In order to validate this equation we assume that the polarizability of the solute molecule does not change discontinuously.

Clearly, evaluation of Eq. [5] requires an expression for the frequency dependent dielectric constant [$\varepsilon(\omega)$] of the solvent continuum. The simplest form of this dielectric constant is given by Debye relaxation,[53] which corresponds to an exponential decay of total dipole moment time correlation function.

$$\varepsilon(z) = \varepsilon_\infty + \frac{\varepsilon_0 - \varepsilon_\infty}{1+z\tau_D} \quad [6]$$

Here, $z = i\omega$, $\tau_D$ is the Debye relaxation time, $\varepsilon_0$ is the static dielectric constant and $\varepsilon_\infty$ is the infinite frequency dielectric constant. By substituting this expression in **Eq. [4]** and performing the required Laplace inversion, we obtain the following expression,

$$r(t) = \left(\frac{3}{\varepsilon_c+2}\right)Fe^{-t/\tau_L} \quad [7]$$

where,

$$F = \left[\frac{2(\varepsilon_0 - \varepsilon_\infty)}{3a^3 \tau_D}\right]\left[\frac{\varepsilon_c + 2}{\varepsilon_c + 2\varepsilon_\infty}\right]^2 \quad [8]$$

and
$$\tau_L = \left[\frac{\varepsilon_c + 2\varepsilon_\infty}{\varepsilon_c + 2\varepsilon_0}\right]\tau_D \quad [9]$$

We note that the longitudinal relaxation time of the reaction field ($\tau_L$) is much smaller than the actual Debye relaxation time ($\tau_D$) for strongly polar liquids such as water. Hence, relaxation of the time-dependent reaction field is exponential when the dielectric relaxation is exponential. This decay is directly compared to the solvation decay in TDFSS experiments.[54]

However, for most complex systems, such as our aqueous DNA system, this simple description is not sufficient. Multiple relaxation processes (nonexponential) are often encountered in experiments and one often considers a summation series[55] such as

$$\varepsilon(z) = \varepsilon_\infty + (\varepsilon_0 - \varepsilon_\infty)\sum_{k=1}^{N}\frac{g_k}{1 + z\tau_k} \quad [10]$$

where, $\sum_{k=1}^{N} g_k = 1$. For example, a bimodal relaxation (also known as Budo formula[55]) is sufficient for polar liquids like ethanol, propanol etc.[56] Following the previous calculations, it can be shown that the decay of reaction field according to Budo relaxation is also biexponential. The most general form of dielectric dispersion, however, is provided by the Havriliak-Negami (HN) equation.[57-58]

$$\varepsilon(z) = \varepsilon_\infty + \frac{\varepsilon_0 - \varepsilon_\infty}{\left[1 + (z\tau)^\alpha\right]^\beta} \quad [11]$$

$\alpha$ and $\beta$ are parameters that lie between 0 and 1. When both these parameters are 1, **Eq. [11]** reduces to the much simpler Debye description in **Eq. [6]**. The analytical inverse Laplace for $r(z)$ using HN relaxation is non-trivial. Hence $r(t)$ needs to be calculated by numerical Laplace inversion.

The above discussion can be extended to include the effects from non-spherical shape of the cavity and also rotation of the solute dipole.[37] In fact, rotation of the probe dipole can be important under certain conditions as it adds an extra channel to the decay. For exponential solvation and single particle rotation for a spherical cavity, the decay constant ($\tau_L$) gets modified to ($\tau_{LR}$)

$$\tau_{LR}^{-1} = \tau_L^{-1} + 2D_R \quad [12]$$

where, $D_R$ is the rotational diffusion coefficient. If we now consider instantaneous change of dipole moment ($\mu$) on excitation, we obtain the following expression for the time dependence of energy shift

$$h\Delta\nu_f(t) = \langle \Delta E(t) \rangle^e - \langle \Delta E(t) \rangle^g = F\tau_{LR}(\mu_g - \mu_e)\left[\mu_e + (\mu_g - \mu_e)e^{-t/\tau_{LR}}\right] \quad [13]$$

Here the superscripts/subscripts '$e$' and '$g$' stand for excited and ground states respectively.

With a non-Debye relaxation of the form presented in **Eq. [11]** for $\alpha = 1$ (also known as the Cole-Davidson relaxation), solvation dynamics follows a stretched exponential form.[54] Consequently, more complex expression for frequency dependent dielectric constant (which is not analytically solvable) needs to be considered in our study.

# V. THEORETICAL ANALYSIS OF THE CONTRIBUTION OF COUNTERIONS TO TIME DEPENDENCE OF SOLVATION ENERGY

The complexity of dynamical relaxation in DNA solutions arises from participation of the several components present in the solution (**Section II**). The role of counterions has not hitherto been studied within a theory. Two phosphate backbones of a DNA duplex can be thought of as polyion chains of equally spaced negatively charged traps (arising from the single bonded oxygen atom) where a fraction of the sites is blocked by the oppositely charged counterions (**Figure 1**). These trapped counterions possess a finite residence time around DNA and move along the backbone from one site to the other, either influenced by the polarization fluctuation around DNA or as a random walker. This phenomenon leads to a fluctuation in the counterion population and generates a current around DNA. In this section we address the plausible origin of slow decay by following a theoretical model proposed by Oosawa.[29]

Oosawa represented the time dependent fluctuations of counterion concentration in terms a diffusion equation in the presence of a force filed due to inter-ionic interaction. When linearized to small fluctuations, the equation admits of a solution by Fourier analysis that leads to a description of fluctuation in terms of a collection of modes, each with a wave number dependent relaxation time. Dielectric relaxation was shown to be related to the decay time of these modes. They contribute significantly to the non-Debye nature of dielectric relaxation. TDFSS solvation time correlation function is related to the frequency dependent dielectric constant by inverse Laplace transform (**Section IV**). Hence, the counterion fluctuations must modify solvation dynamics of a probe bound to DNA. In the following theoretical analysis, we follow the model proposed by Oosawa[29] in order to obtain the contribution and timescales of dielectric relaxation that arise from different modes.

Let $\delta c(x)$ is the spatial fluctuation in concentration of the counterions in the range $x$ to $x+\delta x$. We can express the dipole moment ($\mu$) created by fluctuation of magnitude $\delta c(x)$ as

$$\mu = e\int_0^l \left(x - l/2\right)\delta c(x)dx \qquad [14]$$

where '$l$' is the length of the chain and '$e$' is protonic charge. Now, we expand $\delta c(x)$ in a Fourier series in the following fashion.

$$\delta c(x) = \sum_k \left[a_k \cos(2\pi kx/l) + b_k \sin(2\pi kx/l)\right] \qquad [15]$$

From here on we replace $(2\pi k/l)$ by $k$ for general convenience. Hence, the coefficients of the Fourier series are

$$a_k = \frac{2}{l}\int_0^l \delta c(x)\cos(kx)\,dx; \text{ and } b_k = \frac{2}{l}\int_0^l \delta c(x)\sin(kx)\,dx. \qquad [16]$$

Furthermore, $\langle a_k^2 \rangle = \langle b_k^2 \rangle = \dfrac{1}{\left[(l^2/n)+(l^2/4)\beta\phi_k\right]}$, where $n$ is the average number of counterions bound to phosphate backbones, $\beta$ is $(1/k_B T)$ and $\phi_k$ is the interaction potential between two ions in $k$-space where they are separated by $r$ in the real space. Hence, the expression for mean square fluctuation of induced dipole from **Eq. [14]** becomes

$$\langle \mu^2 \rangle = e^2 l^2 \sum_k \frac{\langle a_k^2 \rangle}{k^2} = 2ne^2 \sum_k \frac{1}{k^2}\left\{\frac{1}{1+(n\beta\phi_k/2)}\right\}. \qquad [17]$$

Therefore, we obtain the polarizability along the phosphate backbone as the sum of polarizabilities of different normal modes of counterion fluctuation.

$$\alpha = \left(\sum_k \alpha_k\right) = \left(\beta\langle\mu^2\rangle_{E=0}\right) = 2n\beta e^2 \sum_k \left(\frac{1}{k^2}\right)\left\{\frac{1}{(1+n\beta\phi_k/2)}\right\}. \quad [18]$$

Now, we follow the procedure described by Onsager[52] to obtain the dielectric increment $(\Delta\varepsilon)$ as a result of the change in polarizability as given in equation [18].

$$\Delta\varepsilon = \varepsilon_0 \left(\frac{Nl}{V}\right)\left(\frac{8\pi}{3}\right) Q \sum_k \frac{1}{k^2}\left(\frac{1}{1+Qw_k}\right) \quad [19]$$

where, $\varepsilon_0$ is the static dielectric constant of the solvent, $\left(\frac{N}{V}\right)$ is the number density of the polyion chain, $Q = \frac{n\beta e^2}{\varepsilon_0 l}$, $w_k = \frac{\varepsilon_0}{e^2}\int_0^l \phi(r)\cos(kr)dr$. The total dielectric increment is also the sum of dielectric increments from different normal modes (i.e., different $k$ values) of counterion fluctuation.

$$\Delta\varepsilon = \sum_k \Delta\varepsilon_k \quad [20]$$

If the counterion concentration ($n/l$) is sufficiently high, it is clear from equation [19] that $\Delta\varepsilon$ becomes independent of $n$. However in our simulation of 38 base pair B-DNA in TIP3P water, we find ~30% sites are populated with $Na^+$ ions whereas the experimentally obtained value is ~50%. In the numerical calculations using equation [19] we assume 50% occupancy. And the contributions from different modes roughly scales are $\frac{1}{k^2}$. The value of $\Delta\varepsilon$ varies from a few hundred to even thousand depending on the number density of DNA molecules. Each mode of relaxation is associated with different relaxation times given by the following expression derived by Oosawa,

$$\tau_k = \left(\frac{1}{Dk^2}\right)\left(\frac{1}{1+Qw_k}\right) \quad [21]$$

For example, we assume a DNA of length 1000 nm that approximately accommodates 3012 base pairs (which indicates 6020 negative charges on the backbone), 50% counterion occupancy (3010 sodium ions along the backbone), $(N/V) \sim 10^{16}/cm^3$, and six modes of counterion concentration fluctuation. The interaction potential energy between ions is $\phi(r) = e^2/r$. We take the diffusion constant as that of the diffusion constant of free sodium ions in water and $\varepsilon_0$ for water is 78 at T=300K. The value for $\Delta\varepsilon$ obtained is 90. We obtain six different timescales from equation [21] that correspond to different modes. These are $\tau_1$=3.5ns, $\tau_2$=880ps, $\tau_3$=396ps, $\tau_4$=225ps, $\tau_5$=145ps and $\tau_6$=101ps. And the contributions of $\Delta\varepsilon$ for different modes are $\Delta\varepsilon_1$= 59.9, $\Delta\varepsilon_2$ = 14.8, $\Delta\varepsilon_3$ = 6.9, $\Delta\varepsilon_4$ = 3.9, $\Delta\varepsilon_5$ = 2.6 and $\Delta\varepsilon_6$ = 1.9. We assume the frequency dependent dielectric response as the sum of six Debye relaxations as follows,

$$\varepsilon(\omega) = \varepsilon_\infty + \sum_{k=1}^{6} \frac{\Delta\varepsilon_k}{1+i\omega\tau_k^D} \quad [22]$$

The Cole-Cole plot obtained by separating the real and imaginary parts of $\varepsilon(\omega)$ shows a marked asymmetry (**Figure 2**). We also plot the contributions of different modes (as single Debye processes) in the same graph to show that $k = 1$ mode is predominant. However, all the modes together gives rise to the non-exponential character of $r(t)$ and asymmetry in the Cole-Cole plot.

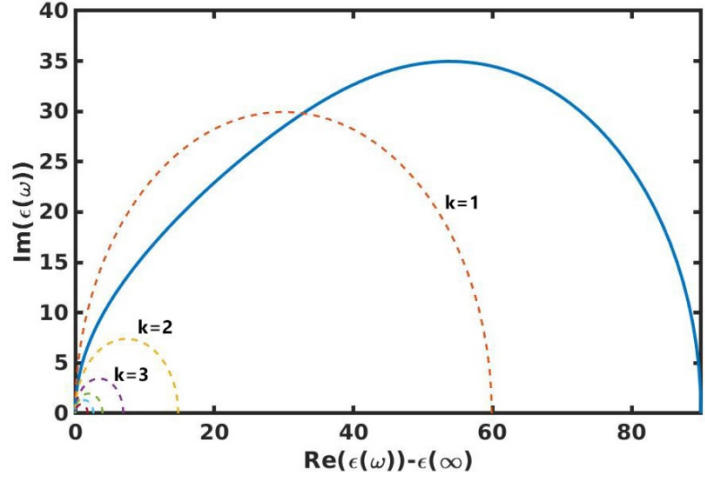

**Figure 2. The Cole-Cole plot obtained by separating the real and imaginary parts of frequency dependent dielectric relaxation as in equation [22] shows a pronounced asymmetry. The solid line (blue) possesses the contribution of all the modes whereas the dashed lines are for individual modes of counterion fluctuation.**

From the expression of frequency dependent permittivity $\varepsilon(\omega)$ we use the formalism developed by Bagchi-Oxtoby-Fleming[37] in order to obtain the solvation time correlation function and corresponding solvation timescales. Bagchi *et al.* showed that the continuum model predicts solvation timescales to be one order of magnitude less that dielectric relaxation timescale. These two are related by **Eq. [9]** in case of a single Debye relaxation process. Typically for neat water at 298K $\tau_D$=8.3 *ps*, which give rise to $\tau_L$ ~500 *fs*. However, this model only works if we assume Debye relaxation process and a spherical probe.

In a related study Bagchi *et al.* showed that a slight modification (for example, Davidson-Cole or Cole-Cole type) in **Eq. [22]** can modulate the solvation timescale to a great extent. We replace $\varepsilon(\omega)$ in **Eq. [4]** by that of **Eq. [22]** to obtain *r*(ω). We perform numerical inverse Laplace transformation by employing Gaver-Stehfest algorithm to obtain *r*(*t*). Now, as discussed earlier, *r*(*t*) is directly comparable to TDFSS time correlation function.

We numerically evaluate and plot [$r(t)/r(0)$] against time using the formalism discussed in **Section IV**, and fit it to two functions as shown in equation [23]- (A) stretched exponential and (B) multiexponential functions. Other functions, for example (i) power-law, and (ii) a summation/product of power law and exponentials do not work out up to the level of satisfaction in this case.

$$\left[\frac{r(t)}{r(0)}\right]_{fitted} = \exp(-t/\tau)^{\beta} \qquad \text{[(A) Stretched exponential]}$$

$$or, = \sum_{i=1}^{3} a_i \exp(-t/\tau_i) \qquad \text{[(B) Tri-exponential]}$$

[23]

We provide the graphs in **Figure 3** and the fitting parameters in **Table 1**.

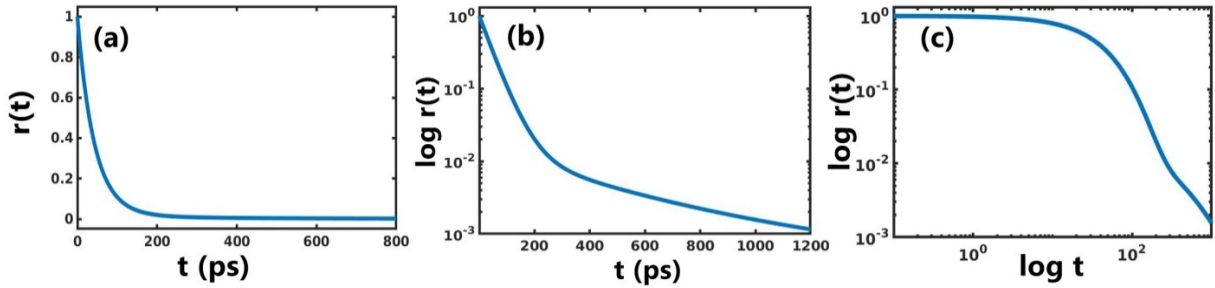

**Figure 3.** (a) Normalized $r(t)$ vs time. (b) Semi-log plot of (*a*) that shows a clear crossover around 300 ps as a trademark of non-exponential character and (c) log-log plot of (*a*).

**Table 1. Fitting parameters of three different plots as shown in equation [23] along with the goodness parameter for thee fittings (namely $\chi^2$ values)**

| Function | $\tau$ (ps) | $\beta$ | $<\tau>$(ps) | | | | |
|---|---|---|---|---|---|---|---|
| (A) | 43.8 | 0.95 | 47.5 | | | | |

| Function | $a_1$ | $\tau_1$ (ps) | $a_2$ | $\tau_2$ (ps) | $a_3$ | $\tau_3$ (ps) | $<\tau>$(ps) |
|---|---|---|---|---|---|---|---|
| (B) | 0.94 | 41.1 | 0.05 | 91.5 | 0.01 | 534 | 48.5 |

If $\varepsilon(\omega)$ is expressed as the sum of several Debye processes *r(t)* becomes,

$$r(t) = \sum_k B(k) \exp\left(-\frac{Dk^2 t}{\tau_k}\right)$$
$$\approx \int_{2\pi/L}^{2\pi/l} dk\, B(k) \exp\left(-\frac{Dk^2 t}{\tau_k}\right) \qquad [24]$$

The integral can be solved as a gamma function integral depending on the nature of coefficient *B(k)*. But, a $\left(1/\sqrt{t}\right)$ factor arises naturally. This contributes to the non-exponentiality and emergence of *R(t)* towards a power law.

## VI. CAN POWER LAW ORIGINATE FROM CORRELATED RANDOM WALK OF IONS?

### a. Continuous time random walk (CTRW) theory

In this section we follow an elegant theoretical description developed by Scher, Montroll and Lax[38-39] in order to explain the anomaly in transit time distribution in photocopy machines and experiments carried out on amorphous solid surfaces. They derived an expression for time dependent current by approximating various forms of hopping time distribution. For our system we found this description to be suitable.

The DNA backbone serves as a chain of negatively charged localized traps where positively charged counterions can reside. There could be three types of transitions (i) counterions arriving from bulk to a site, (ii) ions leaving a site and diffusing to the bulk solvent, and (iii) ions moving along the chain to find another site in the neighborhood. These processes give rise to a distribution of waiting times $[\psi(t)]$. Montroll *et al.* considered the discontinuous flow of ions as propagation of Gaussian current packet. The propagation velocity of the mean position of the packet, $\langle l \rangle$, yields current.

Because of the fluctuation of the counterions, water molecules, and the DNA itself; the probe (located at major/minor groove or intercalated) experiences a continuous fluctuation of the dipolar/ionic electric field. Thus, the electric current $[I(t)]$ generated becomes proportional to the time derivative of the solvation energy [$E(t)$].

$$I(t) \propto \frac{dE(t)}{dt} \qquad [25]$$

Now, in order to move from one site to another is associated with several factors namely a potential barrier, availability of empty sites and separation distance between sites. Moreover, there is spatial dispersion in the location of empty sites and potential barrier that depends on the local surroundings. Hence, one cannot approximate $\psi(\tau)$ as a rapidly decaying function with a single transition rate as often served by an exponential decay. We acquire evidences from our MD simulations that counterions hop along DNA backbone with a *long tailed distribution* of waiting times. Here, we approximate $\psi(\tau)$ as follows,

$$\psi(\tau) \sim \exp[\tau] \int_{\sqrt{\tau}}^{\infty} \left(z - \sqrt{\tau}\right)^2 \exp\left[-z^2\right] dz \qquad [26]$$

such that the long-time tail acquires a $\tau^{-(1+\alpha)}$ dependence; where *k* is the rate constant and $\alpha$ is the exponent between 0 and 1. We now define *P(l,t)* which is the probability of finding a random walker at *l* at time *t*.

$$P(l,t) = \frac{1}{2\pi i} \int_{c-i\infty}^{c+i\infty} dz \frac{\exp(zt)}{z} \left[1 - \psi^*(z)\right] P(l, \psi^*(z)) \qquad [27]$$

where, $\psi^*(z) = \int_0^\infty \exp[-zt]\psi(t)\,dt$ and $P(l,z)$ is the random walk generating function defined as

$$P(l,z) = \frac{1}{N_1 N_2 N_3} \sum_{s_1=1}^{3} \sum_{s_2=1}^{3} \sum_{s_3=1}^{3} \frac{e^{i(\bar{l}.\bar{k})}}{1 - z\lambda(k)} \qquad [28]$$

$k$ is $(2\pi s/N)$ and $\lambda(k)$ is the hopping transition probability in Fourier space expressed as

$$\lambda(k) = \sum_l p(l) e^{i(\bar{l}.\bar{k})}; \; where \; \sum_l p(l) = 1 \qquad [29]$$

Given the probability and waiting time distribution Montroll *et al.* derived that the electric current exhibits power law as follows.

$$\begin{aligned} I(t) &\sim t^{-(1-\alpha)} \; if \; \langle l \rangle \ll L \\ I(t) &\sim t^{-(1+\alpha)} \; if \; \langle l \rangle \geq L \end{aligned} \qquad [30]$$

where $L$ is the total length of the backbone of DNA molecule. Now, from equation [25] it is clear that the solvation energy also exhibits power law either $\sim t^\alpha$ or $\sim t^{-\alpha}$ depending on the two domains mentioned in **Eq. [30]**.

### VII.  ORIGIN OF POWER LAW FROM MODE COUPLING THEORY

As mentioned earlier, several experimental studies on the solvation dynamics of DNA solution bring out the existence of a slow relaxation component (~ ns) that is characterized by a power law. In the previous section we have proposed that the roots of this power law nature lies in the ionic dynamics. In this section we use classical density functional theory (DFT) and mode coupling theory (MCT) to analyze the contribution from ions namely, the Na$^+$ counterions and the ions from buffer solution.

In experiments, solvation dynamics is studied from the time dependent fluorescence Stokes shift (TDFSS) of the emission spectrum of a probe in the solution. The concerned normalized non-equilibrium time correlation function (TCF) is given by **Eq. [1]**. In computational studies, $S(t)$ is translated to a much simpler equilibrium solvation energy time correlation function (TCF) by invoking linear response theory.

$$C(t) = \frac{\langle \delta E(0) \delta E(t) \rangle}{\langle \delta E(0) \delta E(0) \rangle} \qquad [31]$$

In **Eq. [31]**, $\delta E(t)$ denotes the solvation energy fluctuation of the probe at time $t$. According to classical DFT, the time dependent solvation energy of a probe located at position **r** is given by[30, 59]

$$E(\mathbf{r},t) = -\int d\mathbf{r}' \sum_i c_{p,i}(\mathbf{r}') \delta \rho_i(\mathbf{r}-\mathbf{r}',t) - \int d\mathbf{r}' c_{p,w}(\mathbf{r}',\omega) \delta \rho_w(\mathbf{r}-\mathbf{r}',\omega,t) \qquad [32]$$

The first term in **Eq. [32]** arises from the interaction of the probe with positive and negative ions. $c_{p,i}(\mathbf{r})$ is the probe-ion direct correlation function and $\delta \rho_i(\mathbf{r},t)$ is the time dependent number density fluctuation of the ion $i$. The second term comes due to interaction of the probe with water (the terms have similar meaning). Here we have neglected the contributions from DNA as the normal modes of DNA dynamics decay on a much faster time scale.[51]

Now, we take the Fourier transform of the position and time dependent number density and position dependent direct correlation function and define the intermediate scattering function between two ions (*i* and *j*) by the correlation function of their densities in Fourier space (**k** space) (**Eq. [33]**).

$$F_{ij}(\mathbf{k},t) = \frac{1}{N_i N_j} \langle \rho_i(\mathbf{k},t) \rho_j(-\mathbf{k},t) \rangle \qquad [33]$$

Using **Eqs. [32]** and **[33]** we obtain the energy-energy TCF as follows.

$$\langle \delta E(0)\delta E(t) \rangle = \int dk k^2 \sum_{i,j=1}^{2} c_{p,i}(k) c_{p,j}(k) F_{ij}(k,t) \\ + \int dk k^2 [c_{p,w}^2(110;k) F_w(110;k,t) + c_{p,w}^2(111;k) F_w(111;k,t)] \qquad [34]$$

where, $c_{p,i}(k)$ is the probe-ion direct correlation function in **k** space, obtained by Fourier transform of the same in coordinate space, $c_{pw}(11m;k)$ and $F_w(11m;k,t)$ are the spherical harmonic expansions of the probe-water direct correlation function and angle-dependent intermediate scattering function of liquid water respectively.

From computer simulations and NMR experiments we find that the orientational relaxation of water molecules is of the order of ps.[60] Even within DNA grooves, the water molecules exhibit fast rotational dynamics (**Section IX.b.ii**). However, our timescale of interest (that show characteristics of power law) is much longer, often of the order of ns. Therefore, we simplify **Eq. [34]** by dropping the second term in the right hand side, which essentially comes from the fast water orientation. Consequently we are left with the slower dynamics of positive and negative ions. Separating these contributions, we get

$$\langle\delta E(0)\delta E(t)\rangle = A\left[\int_0^\infty d\mathbf{k}k^2 c_{p+}^2(\mathbf{k})F_{++}(\mathbf{k},t) + \int_0^\infty d\mathbf{k}k^2 c_{p-}^2(\mathbf{k})F_{--}(\mathbf{k},t)\right.$$
$$\left. + 2\int_0^\infty d\mathbf{k}k^2 c_{p+}(\mathbf{k})c_{p-}(\mathbf{k})F_{+-}(\mathbf{k},t)\right] \quad [35]$$

Here A is a numerical constant. Our next task is to evaluate the intermediate scattering function of the dynamic structure factor which, according to **Eq. [35]**, determines the behaviour of solvation time correlation function.

Statistical mechanics gives us the following expression of dynamic structure factor in the overdamped limit for monovalent ions (say positively charged).

$$S_{++}(k,z) = \frac{S_{++}(k)}{z + \dfrac{D_+(z)k^2}{S_{++}(k)}} \quad [36]$$

Here, $z = i\omega$ is frequency, $D_+(z)$ is the frequency dependent self-diffusion coefficient of the cation and $S_{++}(k)$ is the partial static structure factor among cations. It is to be noted that **Eq. [36]** is valid when cations and anions are identical in every aspect (except charge) and the solvent is a structureless continuum. This approximation is acceptable as we are primarily interested in the timescale (long) and in the functional form of the decay curve. Detailed calculation involving asymmetric ions (acetate, phosphate etc.) is nontrivial. However, the basic features are easy to capture from this analysis.

$D_+(z)$ is obtained from the generalized Einstein equation given below

$$D_+(z) = \frac{k_B T}{z + \zeta_+(z)} \quad [37]$$

where, $\zeta_+(z)$ is the frequency dependent friction on the cation, which can be decomposed into contributions from Stokes friction and atmospheric electrolyte friction (**Eq. [38]**).

$$\zeta_+(z) = \zeta_{Stokes} + \zeta_{elec}(z) \qquad [38]$$

We have neglected the frequency dependence of Stokes friction due to viscosity.[61] It is to be noted that **Eq. [38]** inherently introduces a timescale separation. While Stokes friction is frequency independent in the range below z ~ $10^{10}$ s$^{-1}$, atmospheric or electrolytic relaxation ($\tau_{atm}$) is often of the order of ns or longer. Now, the electrolytic friction consists of electrophoretic and ion atmosphere terms. However, we may neglect the former as it is significant only at high frequency, which is not of interest in our present problem.

Using statistical mechanics and classical DFT, the frequency dependent electrolytic friction can be expressed as [30, 59]

$$\zeta_{elec}(z) = \frac{q_i^2 \kappa_D}{6\varepsilon D} \frac{1}{1 + \frac{1}{\sqrt{2}}\left[1 + \frac{z}{D\kappa_D^2}\right]^{1/2}} = \frac{\zeta_{elec}(0)\left(1+\sqrt{q}\right)}{1+\sqrt{q(1+z\tau_{atm})}} \qquad [39]$$

Here, $\kappa_D$ is the inverse Debye screening length defined as $\kappa_D^2 = \frac{4\pi\rho q^2}{\varepsilon k_B T}$ and $\tau_{atm}$ is the atmospheric relaxation time given by $\tau_{atm} = \frac{1}{D\kappa_D^2}$. $\zeta_{elec}(0)$ is the zero frequency electrolytic friction and q = ½. It is assumed that all ions have the same diffusion coefficient *D*. From **Eq. [39]** we note that $\zeta_{elec}(z)$ shows a non-exponential temporal decay. **Eq. [39]** needs to be solved self-consistently because of the $\zeta_{elec}(z)$ dependence of *D* on the right hand side.

Therefore from **Eqs. [37]**, **[38]** and **[39]** we get

$$D_+(z) = \frac{k_B T}{z + \zeta_{Stokes} + \dfrac{\zeta_{elec}(0)\left(1+\sqrt{q}\right)}{1+\sqrt{q(1+z\tau_{atm})}}} \qquad [40]$$

It is to be noted here that the frequency dependence of diffusion coefficient is a consequence of the long-range nature of ion-ion interactions. Similar expression is also obtained for negative ions. Now, as mentioned earlier, power law decay is observed in the long time regime (~ ns) of solvation dynamics. The frequency in this scale is small. Considering the range $z\tau_{atm} > 1$, **Eq. [40]** is simplified to

$$D_+(z) \approx \frac{k_B T}{\zeta_{Stokes} + \dfrac{\zeta_{elec}(0)}{\sqrt{z\tau_{atm}}}} \qquad [41]$$

Now, from Stokes-Einstein theory we know

$$D_{Stokes} = \frac{k_B T}{\zeta_{Stokes}} \qquad [42]$$

$D_{Stokes}$ stands for the self-diffusion coefficient for zero ion concentration. This allows us to express **Eq. [41]** in the following form:

$$D_+(z) = \frac{D_{Stokes}\sqrt{z\tau_{atm}}}{\sqrt{z\tau_{atm}} + \dfrac{\zeta_{elec}(0)}{\zeta_{Stokes}}} = \frac{D_{Stokes}\sqrt{\tilde{z}}}{\sqrt{\tilde{z}} + \alpha} \qquad [43]$$

where $\tilde{z} = z\tau_{atm}$ and $\alpha = \zeta_{elec}(0)/\zeta_{Stokes}$. Plugging this into **Eq. [36]**, the expression of dynamic structure factor becomes

$$S_{++}(k,\tilde{z}) = \frac{S_{++}(k)\left(\sqrt{\tilde{z}}+\alpha\right)}{\tilde{z}\left(\sqrt{\tilde{z}}+\alpha\right)+\dfrac{D_{Stokes}k^2}{S_{++}(k)}\sqrt{\tilde{z}}} \qquad [44]$$

Now taking the analytical inverse Laplace transform of **Eq. [44]** in the limit $\alpha > \tilde{z}$ (that is at times longer that atmospheric relaxation) we obtain a power law decay of the scattering function.

$$F_{++}(k,t) = S_{++}(k)\left[e^{\gamma^2 t}\,erfc\left(\gamma\sqrt{t}\right)\right] \qquad [45]$$

where, $\gamma = \dfrac{D_{Stokes}k^2}{\alpha S_{++}(k)}$. When we use this expression of intermediate scattering function in **Eq. [35]**, the power law decay is translated into the relaxation of solvation energy time correlation function. However, one should note that the integrand in **Eq. [35]** contains a $k^2$ term. As a result the contribution from low $k$ modes (where PL is observed) towards solvation relaxation becomes small. *Although the ion atmosphere relaxation gives a power law, the calculated solvation time correlation shows logarithmic dependence, that sets in after 1 ns and contributes approximately 20-30% of the total ion contribution* (*not of the total solvation energy of the whole solution*).[59] This nature is preserved with a slight shortening of logarithmic contribution, even in 1 M solution. We note that the power law nature of $F(\mathbf{k},t)$ is strong when $\alpha = \zeta_{elec}(0)/\zeta_{Stokes}$ is non-negligible. This clearly implies that the power law decay of solvation dynamics at long timescale could arise from a significant contribution of electrolytic friction. Hence a detailed mode-coupling theory is required to establish the microscopic origin of the slow logarithmic/power law decay of solvation dynamics to be from the ionic contribution.

## VIII. EFFECT OF SELF-MOTION OF PROBE ON SOLVATION DYNAMICS

It is important to note that the probe itself can participate in its own solvation process by rotating and translating. An easy way to understand is to think in terms of a free energy surface of solvation that involves solute-solvent interactions, along with solvent-solvent interaction. The Hamiltonian can be written as

$$H = H_P + H_S + V_{PS} \qquad [46]$$

where $H_P$ is the Hamiltonian of the probe, $H_S$ that of solvent and $V_{PS}$ is the probe-solvent interaction. For the simple case where both probe and solvent are ions in a dielectric continuum (the implicit solvent model) we have the following elegant expression for the solvation energy time correlation function of the solvent

$$\langle \delta E_P(0) \delta E_P(t) \rangle = A \sum_{i,j=1}^{\infty} \int_0^{\infty} d\mathbf{k}\, k^2 c_{iP}^2(\mathbf{k}) F_{P,self}(\mathbf{k},t) F_{ij} \qquad [47]$$

where, $i$ and $j$ accounts for both the ionic species of the electrolyte. Note the presence of the self-dynamic structure factor of probe $F_{P,self}(\mathbf{k},t)$. This term decays as $e^{-D_P k^2 t}$ and accelerates the solvation process. Here $D_P$ is the self-diffusion coefficient of the probe. As mentioned above, the motion of the ion facilitates finding the free energy minimum at equilibrium. If the probe is dipolar, the solvation is enhanced by translational motion of the probe, as described in the work of Roy and Bagchi.[62]

We note here that the BOF continuum model can include the rotation of the probe semi-quantitatively but not the translational motion. Rotation of the probe can also contribute to its own solvation process. In the present context, the location of the probe determines whether the self-motion of the probe can have significant contribution. When intercalated inside the double helix with base pair stacking like the original experiment of Berg *et al*. with coumarin dye,[33, 48] the self-motion is totally absent. However, when bound to the groove, the probe can rotate and also translate, to a limited extent. *Thus, when solvation is extra-ordinarily slow, the self-motion of the probe can make a difference. And solvation of the groove bound probe is predicted to be faster than that of the intercalated probe.*

## IX.  ATOMISTIC MOLECULAR DYNAMICS SIMULATION

### a. System and simulation details

We perform atomistic molecular dynamics simulations of DNA-water system using GROMACS[63] package (v5.0.7). The sequence of the 38 base pair DNA has been taken as d(GCCGCGAGGTGTCAGGGATTGCAGCCAGCATCTCGTCG)$_2$, which was earlier shown to produce stable MD trajectories.[64-66] We prepare the initial configuration of the duplex B-DNA using the Nucleic Acid Builder (NAB) module implemented in AMBER.[67] We use amber99sb-ildn force field for DNA[68] and TIP3P water model.[69] Periodic boundary conditions are implemented using a box of dimensions 18 nm × 6 nm × 6 nm filled with 20424 water molecules and 74 Na$^+$ ions, maintaining the minimum image condition. We perform energy minimization of the whole system using steepest descent algorithm followed by conjugate gradient method. Thereafter, the system is subjected to simulated annealing[70] in order to heat it up from 300K to 320K and again cool it down to 300K to help it to get out of a local minima (if any). The solvent is then equilibrated for 5 ns at constant temperature (300 K) and pressure (1 bar) (NPT) by restraining the positions of the DNA atoms followed by NPT equilibration for another 5 ns without position restrains. The final production runs have been carried out at a constant temperature (T=300 K) (NVT) for 25 ns. For analyses we use the last 20 ns of the trajectory to avoid effects of the removal of barostat. The equations of motions are integrated using leap-frog integrator with an MD time step of 2 fs (also the data dumping rate). We have used modified Berendsen thermostat[71] ($\tau_T$ = 0.1 ps) and Parrinello-Rahman barostat[72] ($\tau_P$ = 2.0 ps) to maintain constant temperature and pressure respectively. The cut-off radius for neighbour searching and non-bonded interactions are taken to be 10 Å and all the bonds have been constrained using the LINCS[73] algorithm. For the calculation of electrostatic interactions, we have used Particle Mesh Ewald (PME)[74] with FFT grid spacing of 1.6 Å. All reported data are averaged over three MD trajectories starting from entirely different configuration of the system.

**b. Dynamics of DNA hydration layer and groove water molecules**

**i. Selection of Groove Water Molecules**

In order to calculate groove specific properties and relaxation of water, we precisely select major and minor groove water molecules residing in DNA hydration layer. First we obtain the radial distribution function between carbon atoms specific to major / minor groove and water molecules. The first minimum is found to be around 0.45 nm. Hence, we take the cut-off distance to be 0.5 nm. Additionally, we apply an angle cut-off to distinguish groove water molecules – the angle between particular DNA body-fixed vectors (shown by white arrows in **Figure 4**) and a vector that connects carbon atoms (specific to a particular groove) and oxygen of water molecules (shown by green arrows in **Figure 4**). The body-fixed vectors points outside and perpendicular to an imaginary surface that spirals through two different grooves. This vector continuously changes direction as one progress along the helix. **Figure 4** shows a snapshot from the simulation (at time = 0) where the major and minor groove water molecules are shown in different colours. We calculate the residence time of these water molecules in the respective grooves and consider only those which reside more than 50 ps in order to obtain statistically significant time correlation functions. At the initial frame we find 143 minor groove water molecules and 192 major groove water molecules. Furthermore, we shift the time origin in every 5 ns to take into account groove water molecules.

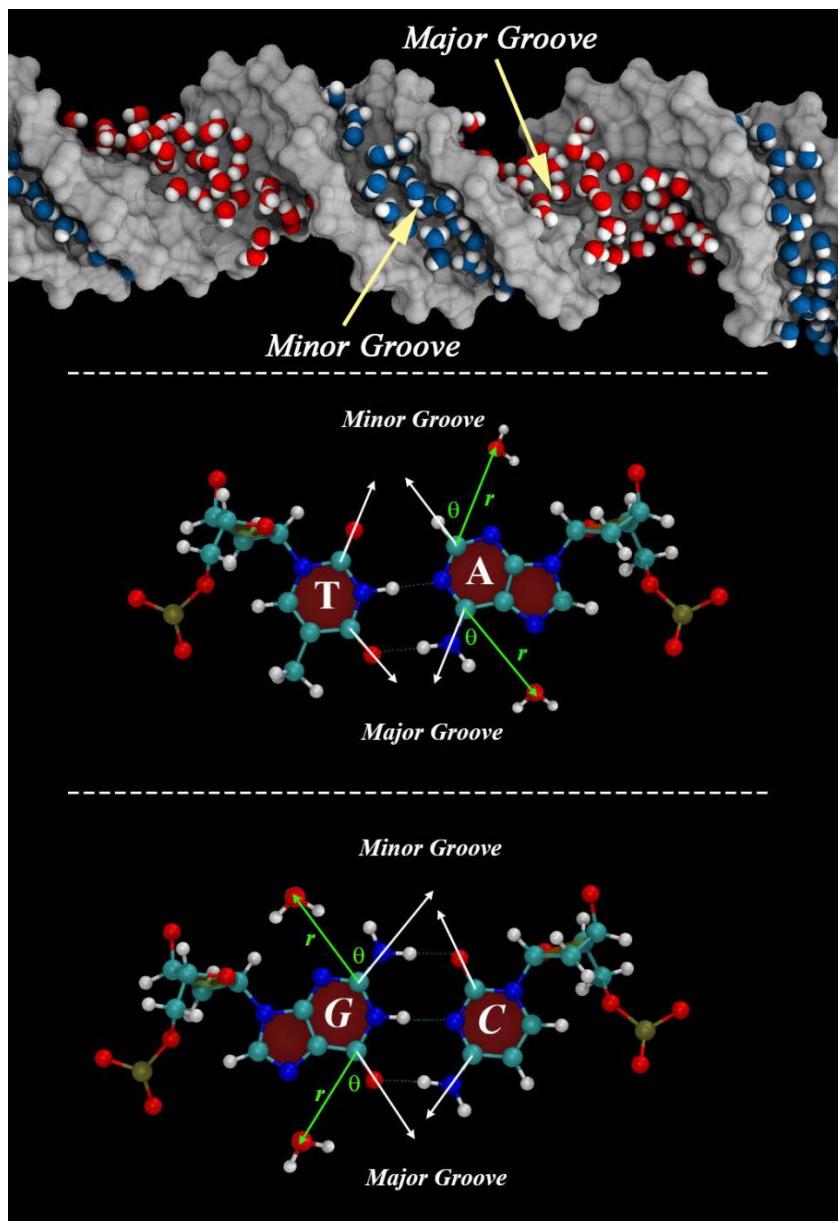

**Figure 4. An illustration of the system and water selection scheme. We explicitly show major and minor groove hydration layer water molecules in different colours. The water molecules that are shown in red belong to major groove whereas those shown in blue belong to minor groove. We also show two representative Watson-Crick base pairs (AT and GC) and how the water selection is performed. The white arrows indicate DNA body fixed vectors and green arrows indicate vectors connecting the carbon atoms with oxygen atom of water. A water molecule is selected as a particular groove water if $0 \leq r \leq 0.5\,\text{nm}$ and $-\pi/2 \leq \theta \leq \pi/2$.**

## ii. Orientational Correlation

We aim to distinguish the DNA hydration layer from the bulk by calculating single particle first and second rank rotational time correlation functions (**Eqs. [48] and [49]**) of the water molecules that reside more than 50 ps in DNA minor and major grooves.

$$C_1^i(t) = \left\langle P_1\left(\hat{\boldsymbol{\mu}}^i(0)\cdot\hat{\boldsymbol{\mu}}^i(t)\right)\right\rangle; \quad P_1(\cos\theta) = \cos\theta \quad [48]$$

$$C_2^i(t) = \left\langle P_2\left(\hat{\boldsymbol{\mu}}^i(0)\cdot\hat{\boldsymbol{\mu}}^i(t)\right)\right\rangle; \quad P_2(\cos\theta) = \tfrac{1}{2}\left(3\cos^2\theta - 1\right) \quad [49]$$

Here, '$i$' denotes the water index in the groove, $P_1$ and $P_2$ are respectively the first rank (associated with optical Kerr effect) and second rank (associate with polarization anisotropy) Legendre polynomials. We obtain both the time correlation functions for each water molecule and fit to a biexponential function using '*lsqcurvefit*' routine implemented in MATLAB. We calculate the individual rotational time constants by estimating the area under each biexponential curve as shown in **Eqs. [50]**.

$$\left\langle \tau_k^{(i)} \right\rangle = \int_0^\infty C_k^i(t)\,dt \quad (k=1,2) \quad [50]$$

We plot the distributions of $\left\langle \tau_1^{(i)} \right\rangle$ and $\left\langle \tau_2^{(i)} \right\rangle$ along with the same quantities in neat water. **Figure 5a** and **Figure 5b** show the distributions of first rank rotational constants and **Figure 5c** and **Figure 5d** show the same for second rank rotational constants. We notice that the relaxation time constants exhibit an unusually broad log-normal distribution with a small fraction of water molecules that are faster than bulk and a long tail extending up to ~100 ps. The distribution is narrow in the bulk. In comparison to the protein hydration layer, the distribution is less extended. This can be rationalized by considering the different bio-molecular surfaces. Proteins are made up of large number peptides (each peptide could be one of the permutation of 20 types of amino acids) whereas for DNA the choices are less. Hence, the degree of surface heterogeneity is less and there are certain self-similarities among the regions on the surface.

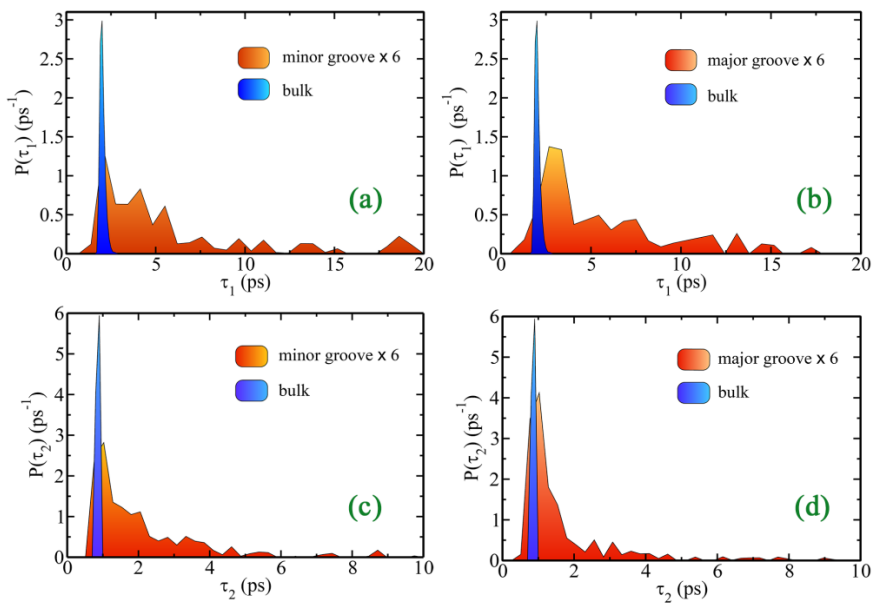

**Figure 5.** Distribution (red) of (a) first rank rotational time constants ($\tau_1$) for minor groove hydration layer, (b) first rank rotational time constants ($\tau_1$) for major groove hydration layer, (c) second rank rotational time constants ($\tau_2$) for minor groove hydration layer and (d) second rank rotational time constants ($\tau_2$) for major groove hydration layer. All the four distributions are compared to corresponding bulk distributions (blue). In all the four scenarios the dynamics is heterogeneous in the DNA hydration layer with a higher fraction of slowly rotating and a lower fraction of faster (or almost as fast as bulk) rotating water molecules. Also, the distribution in minor groove is broader than that of major groove.

It is also clear from **Figure 5** that the distributions in case of minor groove water molecules are broader compared to major groove water molecules implying higher dynamical heterogeneity in minor grooves. The distribution can be directly correlated with the heterogeneous solvation dynamics that is often observed in minor/major groove.[17] If the probe is placed in a region where water molecules exhibit bulk like or even faster rotation, solvation is faster.

In **Figure 6** we show the particle averaged first and second rank rotational time correlation functions – the ones often measured by NMR experiments. Here, in accordance with the previously reported experiments, we find that groove water molecules are only ~2-5 times slower than the bulk. Furthermore, minor groove water molecules are approximately 1.5 times slower compared to major groove water molecules. This happens because minor grooves from deep pockets whereas major groove waters are more exposed. We fit the resulting correlation functions to multi exponential functions and list the timescales in **Table 2**. We find the timescales arising from the groove water molecules are almost one order of magnitude higher compared to that of the bulk.

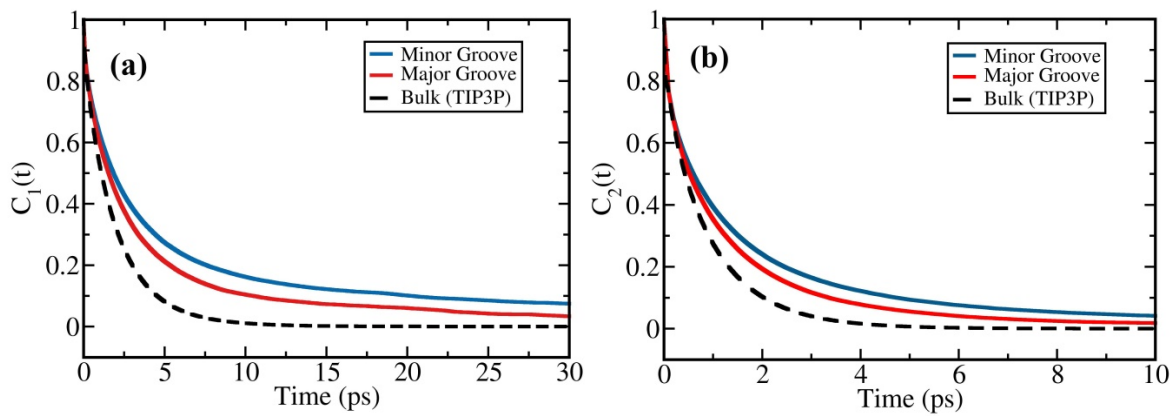

**Figure 6. Normalized and particle averaged (a) first rank and (b) second rank rotational time correlation function for bulk (neat TIP3P, black dashed line), minor groove (blue line) and major groove (red line) hydration layers. There is a noticeable difference in the relaxation process that is distinct from bulk by time constants at least one order of magnitude higher. Moreover the relaxation in minor groove hydration layer is approximately 1.5 times slower compared to that of the major groove which provides evidences in favour of a more restricted environment in minor groove.**

**Table 2. Multi-exponential fitting parameters for first and second rank rotational correlation of bulk, DNA minor groove and major groove water molecules at 300K.**

| | \multicolumn{6}{c}{**First rank rotational correlation function [$C_1(t)$]**} |
|---|---|---|---|---|---|---|
| | $a_1$ | $\tau_1$ (ps) | $a_2$ | $\tau_2$ (ps) | $<\tau>$ (ps) | Average Retardation |
| **Minor Groove** | 0.72 | 1.71 | 0.28 | 20.52 | 6.92 | 3.8 |
| **Major Groove** | 0.76 | 1.60 | 0.24 | 14.01 | 4.51 | 2.5 |
| **Bulk** | 0.16 | 0.15 | 0.84 | 2.12 | 1.80 | --- |

| (TIP3P) | | | | | | |
|---|---|---|---|---|---|---|
| Second rank rotational correlation function [$C_2(t)$] | | | | | | |
| | $a_1$ | $\tau_1$ (ps) | $a_2$ | $\tau_2$ (ps) | $<\tau>$ (ps) | Average Retardation |
| **Minor Groove** | 0.70 | 0.56 | 0.30 | 5.16 | 1.94 | 2.5 |
| **Major Groove** | 0.48 | 0.26 | 0.52 | 2.17 | 1.25 | 1.6 |
| **Bulk (TIP3P)** | 0.21 | 0.04 | 0.79 | 0.97 | 0.77 | --- |

### iii. Translational Diffusivity and Velocity Autocorrelation Function

We estimate the heterogeneity in translational motion of groove water molecules by plotting the distribution of time required to diffuse one molecular diameter (0.315 nm for TIP3P water) in the groove and the neat water. The broad distribution is also present here in case of groove water molecules. In neat water the distribution is sharp. However there are not many differences in the distributions for major and minor groove water molecules.

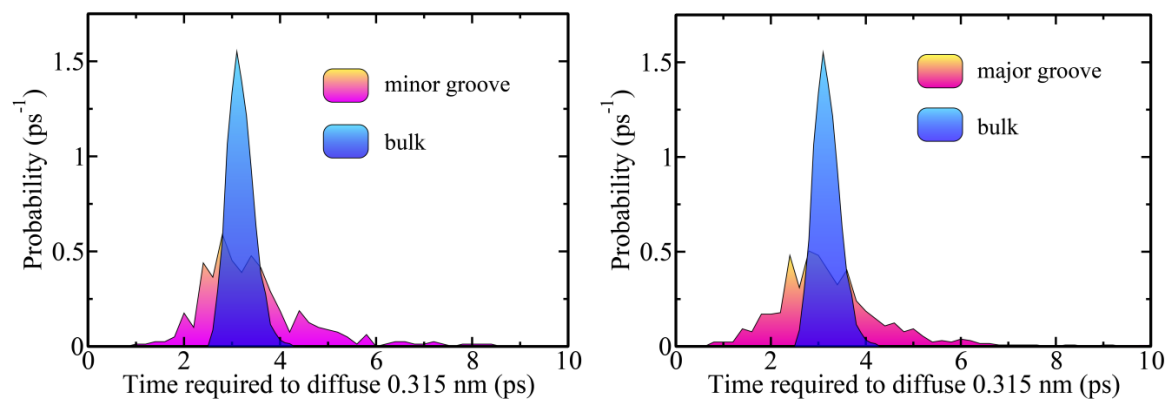

**Figure 7.** Probability of time required to diffuse one molecular diameter (0.315 nm for TIP3P water model) for (a) DNA minor groove hydration layer and (b) DNA major groove hydration layer compared to the same in the bulk (shown in blue). It is clearly seen that the distribution for the hydration layer is broader compared to bulk but the span is almost similar in right and left had side of the distribution of the bulk. This gives the average value approximately equal to the bulk.

In order to differentiate between the translational dynamics of these two regions we calculate translational velocity autocorrelation functions (VACF) of water molecules (by considering velocities of oxygen atoms). In **Figure 8** we plot the normalized VACF for individual water molecules in the grooves (**Figure 8a** and **Figure 8b**) and the bulk (**Figure 8c**) as well as the particle averaged VACF (**Figure 8d**). It is again clear that the groove water molecules show heterogeneous translational motion and a spectrum of relaxation in the VACFs unlike neat water. Hence, the distribution is omnipresent.

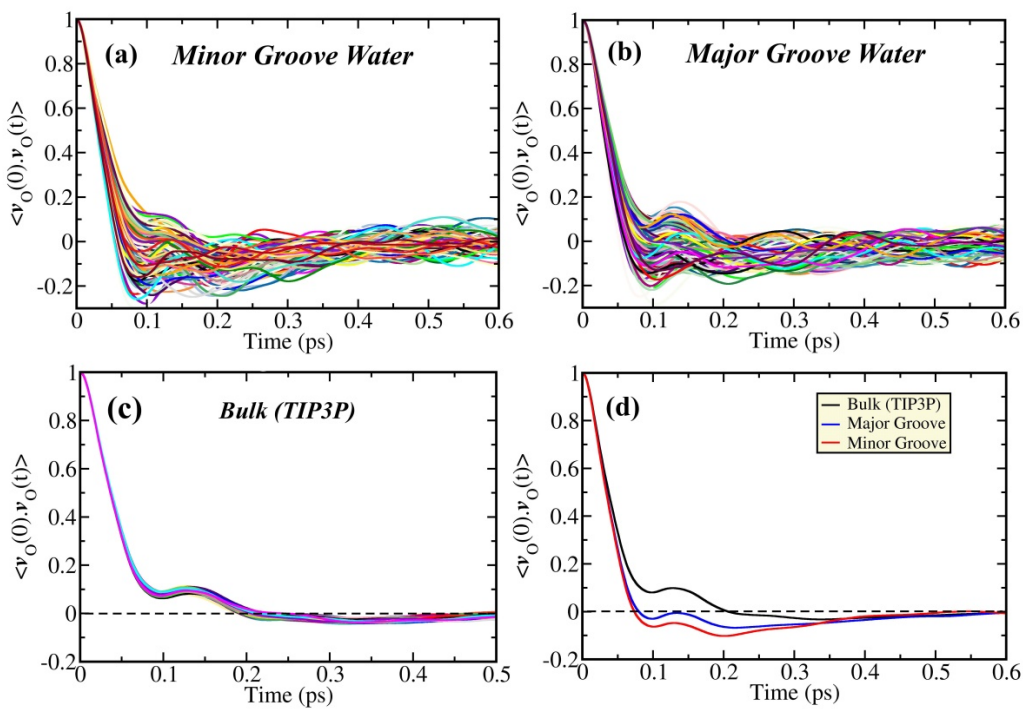

**Figure 8. Single particle and averaged translational velocity autocorrelation function (VACF) of the bulk and hydration layer plotted against time. (a) Single particle translational velocity autocorrelations for DNA minor groove hydration layer water molecules, (b) Single particle translational velocity autocorrelations for DNA major groove hydration layer water molecules, (c) Single particle translational velocity autocorrelation for water molecules in the bulk (TIP3P), and (d) particle averaged translational VACF function for then bulk (black), DNA major groove (blue) and minor groove (red) water molecules. The individual autocorrelations exhibit a spectrum of relaxation in the hydration layer in contrast with the bulk where all water molecules show similar trend of relaxation. The particle averaged VACFs for groove water molecules exhibit marked difference from the bulk and supports a constrained translation there.**

The particle averaged VACF shows an expected constrained translation in DNA grooves, minor groove water molecules being more constrained than those of major groove. We use the translational velocity autocorrelation functions to obtain an estimate of the diffusion constants in minor and major grooves by the use of well-known Green-Kubo formalism shown in Eq.[51]. The results are shown in **Table 3**.

$$D = \frac{1}{3}\int_0^\infty \langle v_O(0) \cdot v_O(t) \rangle dt \quad [51]$$

**Table 3. Coefficient of diffusion as obtained by integrating the unnormalized translational velocity autocorrelation functions. The normalized forms are shown in Figure 8d. The data show that water molecules in minor grooves are almost 1.6 times translationally more constrained than those of major groove water molecules.**

| Water type | Diffusion Coefficient ($cm^2/s$) |
|---|---|
| Minor Groove | $1.56 \times 10^{-5}$ |
| Major Groove | $2.61 \times 10^{-5}$ |
| Bulk (TIP3P) | $5.92 \times 10^{-5}$ |

Diffusion coefficients are directly related to configurational entropy by scaling relations like Rosenfeld scaling or Adam-Gibbs relation.[40]

### c. Solvation dynamics

### (i) Solvation dynamics of metformin

As mentioned in **Section VII**, computationally it is easier and accurate to study the solvation dynamics from the TCF of interaction energy of the probe with the rest of the system (**Eq. [31]**) (under linear response approximation). Here we study the TCF relaxation of a small molecule metformin that is bound to the minor groove of DNA. For our investigations, we have chosen two kinds of DNA: AT-rich (Adenine-Thymine base pair) and GC-rich (Guanine-Cytosine base pair). The results are presented in **Figure 9**.

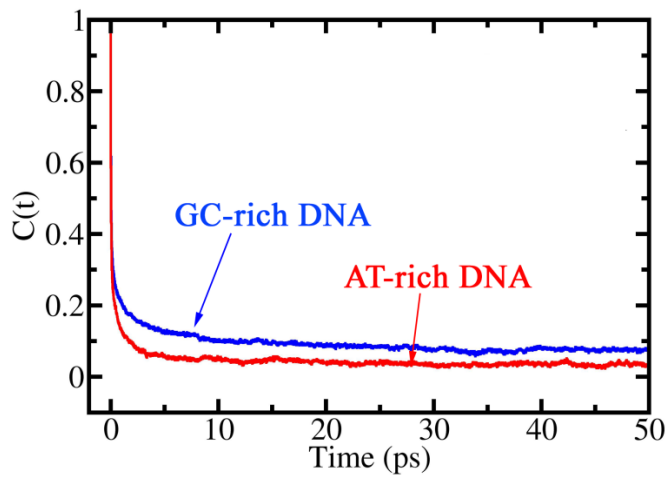

**Figure 9. Solvation dynamics of metformin bound to the minor grooves of AT-rich and GC-rich DNA. Relaxation is faster when the DNA is rich in GC base pairs. The relaxations do not show signature of power law in the timescale of our simulation.**

We fit the above time correlation functions to several functions, such as multiexponential, stretched exponential and power law. Among these, triexponential fitting provides the best correlation with data. The fitting parameters are given in **Table 4**.

$$C(t) = \sum_{i=1}^{3} a_i e^{-\frac{t}{\tau_i}} \qquad [52]$$

**Table 4. Triexponential fitting parameters of solvation dynamics relaxation of metformin bound to minor groove of DNA.**

| DNA Type | $a_1$ | $\tau_1$ (ps) | $a_2$ | $\tau_2$ (ps) | $a_3$ | $\tau_3$ (ps) | $\langle\tau\rangle$ (ps) |
|---|---|---|---|---|---|---|---|
| AT rich | 0.84 | 0.215 | 0.11 | 2.32 | 0.05 | 224.0 | 11.64 |

| | | | | | | |
|---|---|---|---|---|---|---|
| **GC rich** | 0.80 | 0.253 | 0.11 | 4.83 | 0.09 | 172.96 | 16.30 |

From **Table 4** we see that relaxation is slower in case of GC rich DNA. We do not observe any power law decay in solvation dynamics as the simulations are relative shorter (5 ns) than the timescales at which such nature is observed in experiments (~100 ns).

**(ii)    Solvation dynamics of DNA bases**

In experiments, one often studies the solvation dynamics of mutated bases, for example amino purine.[24, 34] In our simulation, we have studied the solvation energy TCF of four different types of bases (Adenine [A], Thymine [T], Guanine [G] and Cytosine [C]) present in the DNA. The results reported here are from a 5 ns simulation of a 38 base-pair long DNA.

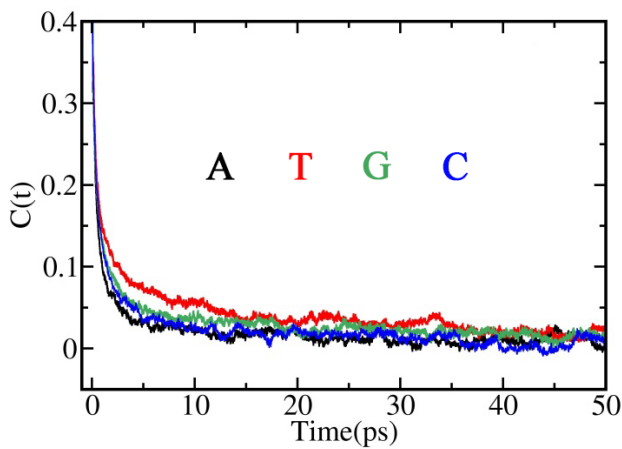

**Figure 10. Solvation dynamics of 4 bases (A, T, G and C) in a 38 base pair DNA. (a) linear scale and (b) log-log scale. The relaxations do not show signature of power law in the timescale of our simulation.**

**Figure 10** shows the relaxation of solvation dynamics of the four bases. In this case as well the fitting is best represented by a triexponential function. The fitting parameters are presented in **Table 5**.

**Table 5. Triexponential fitting parameters of solvation dynamics of four bases in DNA.**

| Base | $a_1$ | $\tau_1$ (ps) | $a_2$ | $\tau_2$ (ps) | $a_3$ | $\tau_3$ (ps) | $\langle\tau\rangle$ (ps) |
|---|---|---|---|---|---|---|---|
| A | 0.70 | 0.039 | 0.27 | 0.94 | 0.03 | 56.4 | 1.97 |
| T | 0.74 | 0.052 | 0.20 | 1.89 | 0.06 | 60.5 | 4.05 |
| G | 0.74 | 0.047 | 0.22 | 1.53 | 0.04 | 112.8 | 4.88 |
| C | 0.69 | 0.040 | 0.26 | 1.04 | 0.05 | 27.05 | 1.65 |

Hence, we find that the relaxations of the bases are faster than a groove bound probe. The difference in timescales can arise from a variety of sources including effects from DNA dynamics, water, ions etc. Further investigations are required to clearly understand the timescales.

### d. Ion hopping along phosphate backbone

From our molecular dynamics simulations we find strong evidence of ions hopping along and across the two backbones of the DNA duplex. In **Figure 11** we represent hopping trajectories of four such Na$^+$ counterions. These ions not only travel along a single DNA strand, but are also found to execute inter-strand jumps.

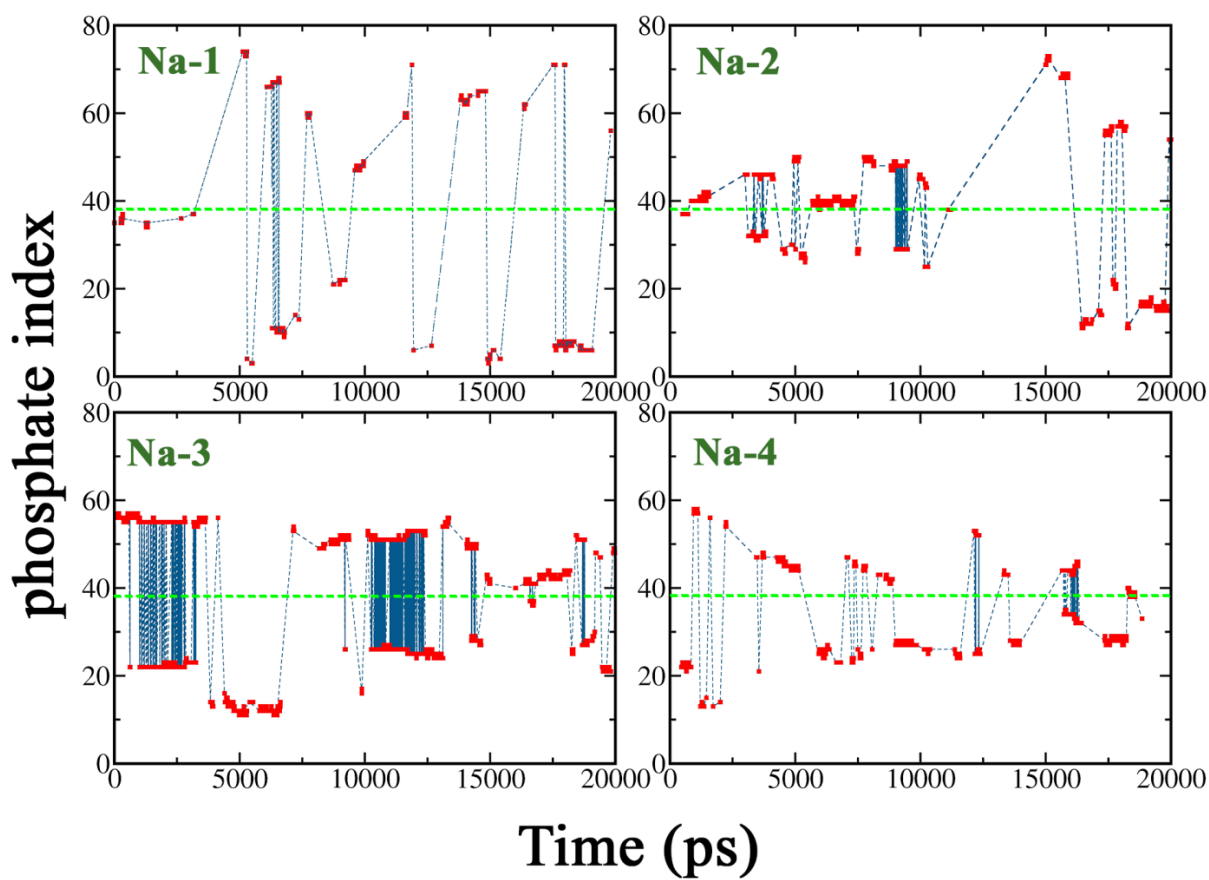

**Figure 11.** Hopping of Na$^+$ counterions along phosphate backbone in a DNA duplex obtained from atomistic molecular dynamics simulations with explicit water. Figures are shown for four such ions. This phenomenon is observed for other Na$^+$ ions as well. From our simulation we observe both intra-chain and inter-chain hopping (large amplitude jumps in the figure).

In the simulated DNA duplex there are 38 base pairs, corresponding to a total of 76 phosphate groups in the two strands. In the above figure, the phosphate groups are numbered 1 to 38 along one strand (marked with green dashed lines in **Figure 11**) and 39 to 76 along the other in the opposite direction. Consequently the serial numbers of two nearest phosphate groups in two different chains are not consecutive. Hence if an ion hops between two such groups, we get a long amplitude jump. This scheme helps us to distinguish between inter-chain and intra-chain counterion jumps.

Thus the dynamics of these ions considerably modify the polarizability and the dielectric properties of the environment around the probe. Consequently, the solvation dynamics is significantly perturbed by this ionic contribution.

## X.    KINETIC MONTE CARLO SIMULATION

From theoretical calculations we have shown that counterions significantly contribute to the solvation dynamics of an aqueous DNA solution. In this section we propose a model system to understand the dynamics of these ions. We model the $Na^+$ counterions as 1-dimensional random walkers moving along the DNA phosphate backbone which is approximated as an energy landscape with the phosphate groups located at the energy minima. $Na^+$ ions can get captured in these energy minima due to attractive interactions with the negatively charged phosphate groups. We consider $N$ such phosphate sites corresponding to the number of base moieties present in the DNA.

In this model, we approximate that a counterion present at site $i$ can hop on to site ($i+1$) or ($i-1$) if either of the latter is vacant. In doing so, the ion has to overcome an energy barrier that separates the two minima (due to attraction from phosphate). For simplicity we neglect the jumps from $i$ to ($i \pm n$) (for $n > 1$) and vice versa as their probabilities are negligible. This movement of ions along the phosphate backbone is controlled by a probability factor that is biased towards the probe used for solvation dynamics studies.

In our model we also consider diffusion of ions from bulk to the backbone and vice versa. We note that adsorption of ions on the phosphate sites is dependent on the concentration of the ions in the immediate vicinity of the site. The model is summarized in **Figure 12**.

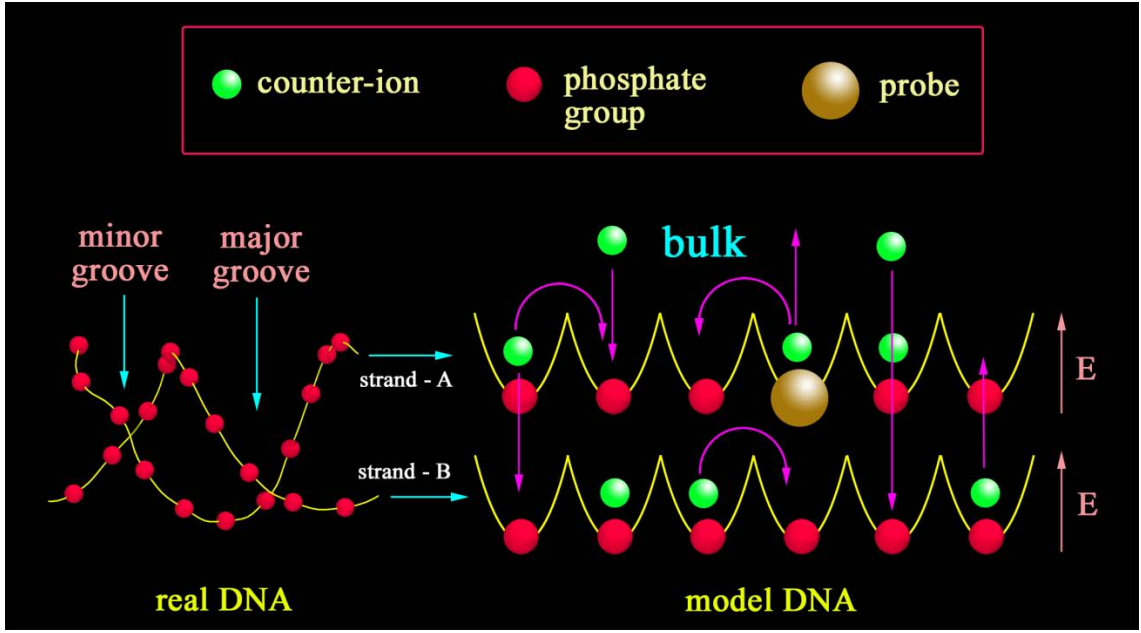

**Figure 12. Schematic representation of the model considered in this work for studying the dynamics of counterions along DNA Phosphate backbone. Some of the possible counterion movements are denoted by magenta arrows.**

Considering all these factors, the dynamics of the random walkers can be described by the following master equation:

$$\begin{aligned}\frac{dP_i}{dt} = &-k_{i\to(i+1)}P_i\Theta_i\{1-\Theta_{(i+1)}\}+k_{(i+1)\to i}P_i\Theta_{(i+1)}\{1-\Theta_i\} \\ &-k_{i\to(i-1)}P_i\Theta_i\{1-\Theta_{(i-1)}\}+k_{(i-1)\to i}P_i\Theta_{(i-1)}\{1-\Theta_i\} \\ &-k_{i\to B}P_i\Theta_i+k_{B\to i}P_i C_B^i\{1-\Theta_i\}\end{aligned} \qquad [53]$$

In the above equation, $P_i$ is the probability density of counterions at site $i$. $k_{i\to j}$ denotes the rate of movement of an ion from site $i$ to site $j$. In this context, "$B$" stands for bulk. $C_B^i$ is the concentration of ions in the bulk near site $i$. In Eq. [53] we introduce an occupancy parameter $\Theta_i$ which defines whether the site $i$ is occupied or vacant. Vacancy is denoted by '0' whereas its value is '1' f the site is occupied.

A few comments regarding Eq. [53] are in order. The first term in the equation describes the movement of ions from a site $i$ to the adjacent site ($i+1$). The second term denotes the movement in opposite direction. The third and the fourth terms describe similar movements between sites $i$ and ($i$-1). The last two terms stand for movement of ions to and from the bulk respectively.

While analytical solution of this equation is non-trivial, the motion of ions, particularly in presence of a probe, can be studied from a Monte Carlo simulation. We perform such a simulation based on the above mentioned principles. We consider a system of 100 phosphate sites and simulate the system with a probability of 0.6 for movement of counterions towards the probe (considering it to have charge density greater than the phosphate groups). The probe is modeled to be located after the $100^{th}$ site. The energy barrier between two phosphate minima is taken to be 1.5 $k_B T$. We run the simulation for $1\times10^5$ Monte Carlo (MC) steps and average the data over 1000 initial configurations. Initial occupancy in each case was taken to be 50 %.

Because of the bias, the ions tend to queue towards the probe by the end of the simulation. This reflects the situation where ionic contributions can largely affect the environment of the probe. The queuing becomes evident when we look and the occupancy distribution of the phosphate sites. Due to this accumulation the distribution is skewed towards the probe (**Figure 13**).

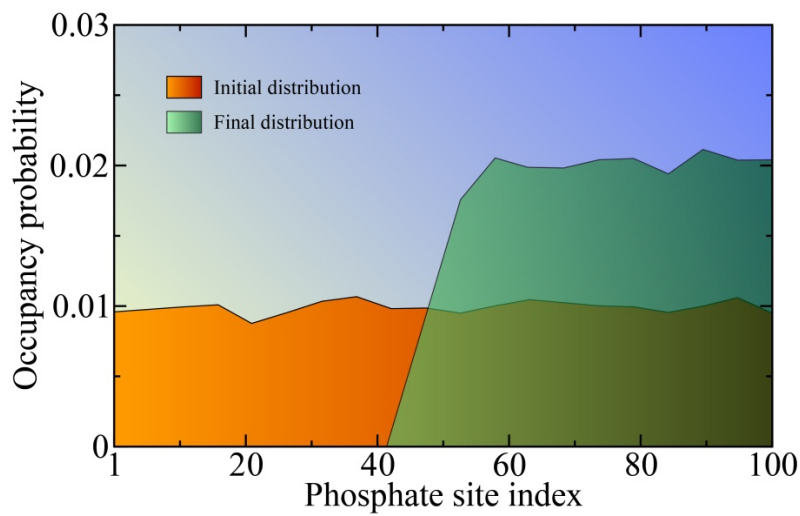

**Figure 13. Counter ion occupancy probability distribution of phosphate sites before and after simulation. Distribution gets skewed towards the probe (after the 100$^{th}$ site) due to bias introduced for movement of ions.**

From the simulation we calculate the total interaction energy between the ions present in different minima and the probe (with a dipolar electric field) at every MC step. We then calculate the energy time correlation function, considering that each MC step corresponds to time separated by a constant time gap. This shows a non-exponential decay as shown in **Figure 14**.

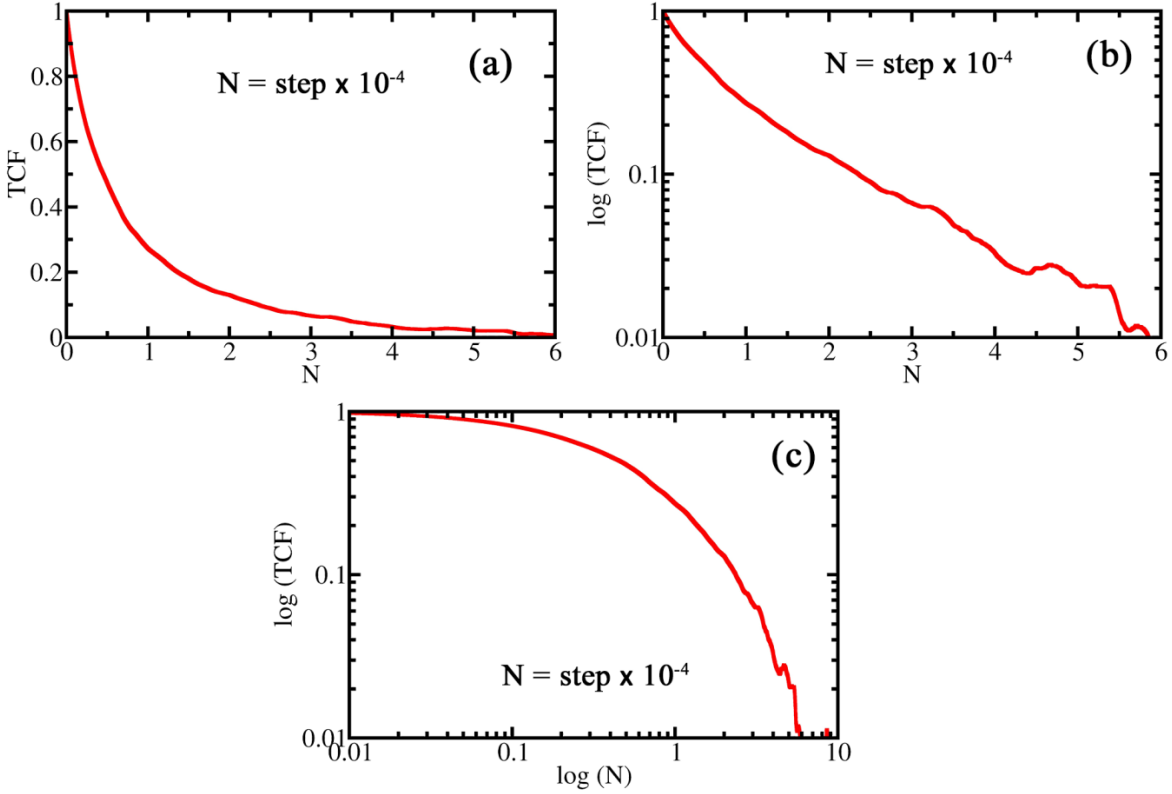

**Figure 14.** Non-exponential decay of the normalized time correlation function of interaction energy between ions present on the model phosphate backbone and the probe. Figure shows representation in (a) linear, (b) semi-log and (c) log-log scales.

The relaxation of the energy TCF shows nonexponential behaviour. We fit the above graph with two different nonexponential functions: (a) triexponential and (b) power law multiplied by a biexponential. Both the fits show high correlation with data. The fitting parameters are shown in **Table 6**.

**Table 6. Fitting parameters of energy TCF for two functional forms.**

| Triexponential: $C(t) = \sum_{i=1}^{3} a_i e^{-t/\tau_i}$ ||||||| 
|---|---|---|---|---|---|---|
| $a_1$ | $\tau_1 \times 10^4$ | $a_2$ | $\tau_2 \times 10^4$ | $a_3$ | $\tau_3 \times 10^4$ | $\langle \tau \rangle \times 10^4$ |
| 0.08 | 0.08 | 0.52 | 0.45 | 0.4 | 1.67 | 0.91 |
| Power law multiplied to biexponential: $C(t) = \left(1 + \dfrac{t}{\tau_0}\right)^{-\alpha} \times \sum_{i=1}^{2} a_i e^{-t/\tau_i}$ |||||||

| $\tau_0 \times 10^4$ | $\alpha$ | $a_1$ | $\tau_1 \times 10^4$ | $a_2$ | $\tau_2 \times 10^4$ |
|---|---|---|---|---|---|
| 0.09 | 0.11 | 0.49 | 0.52 | 0.51 | 1.81 |

The time correlation function showed above is the solvation dynamics relaxation of the probe due to the ionic contribution. The nature of this decay further substantiates the inference that the nonexponential relaxation of solvation dynamics in DNA solution might have its roots in the ionic environment.

## XI.  CONCLUSIONS

In this paper we present a combined analytical and computer simulation study of DNA solvation dynamics in aqueous solution. This study brings out a number of novel aspects. More importantly, it helps to generate a unified picture of the temporal stages of solvation dynamics. This understanding is schematically represented in **Figure 15**.

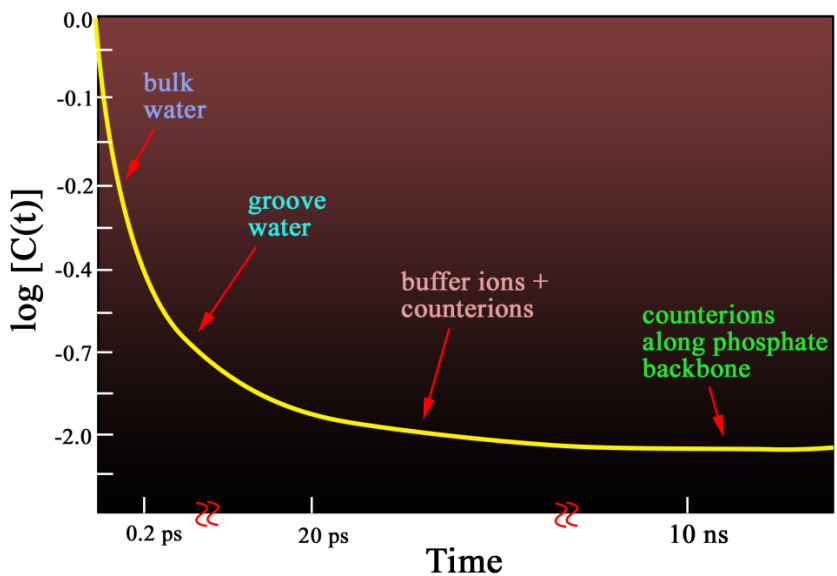

**Figure 15. Schematic representation of solvation dynamics in aqueous DNA solution. The relaxation is characterized by multiple time scales ranging from fs to ns. The primary participants in this dynamics are water (bulk and groove bound) and ions (counterions and those from buffer solution). While water contributes to the initial fast decay, movement of counterions along the phosphate backbone accounts for the slow relaxation in the ns time regime.**

The main results of the present work can be summarized as follows.

1. We have used counterion fluctuation theory of polyelectrolyte dynamics[29] to investigate the effects of counterions on dielectric relaxation and solvation dynamics. The theory produces a pronounced non-Debye form of dielectric relaxation. When we directly include the contribution of the counterions to the solvation dynamics of an ionic probe, we find a power law decay of the form $1/\sqrt{t}$.

2. However, when the composite dielectric function is used in the dielectric continuum model of Bagchi, Fleming and Oxtoby (BOF),[37] the slow decay due to the counterions becomes much weaker. This clearly marks a failure of the dielectric continuum model to describe individual components when a composite dielectric function is employed.

3. Taking forward from Oosawa theory, we employed a continuous time random walk (CTRW) model of Scher, Montroll and Lax[38-39] to study the effects of correlated random walk of counterions on the phosphate backbone, on the solvation dynamics of a polar probe. The theoretical analysis produces a temporal power law.

4. Since a detailed analytical solution of the CTRW model is difficult, we carried out a kinetic Monte Carlo simulation of the random walk and calculated the solvation time correlation function. The calculation also produces a slow nonexponential decay. The nature of the decay was found to depend on various parameters of the model.

5. In order to check the viability of the Oosawa-Scherr-Montroll-Lax model of counterion motion, we carried out explicit MD simulations where we tagged the motion of many sodium ions. We found that the counterions indeed hop between different phosphate sites and such motions can be expressed as a coarse-grained random walk.

6. We revisited the mode coupling theory (MCT) analysis of solvation dynamics of a probe by the ions of an electrolyte solution. MCT predicts power law decay. The analysis is complicated for unequal sized ions, with the slow time determined partly by the diffusion constant of the larger of the two ions (acetate or phosphate).

7. We carried out atomistic molecular dynamics simulations upto 20 ns with a system consisting either of a dodecamer or a DNA with 38 base pair. We explore the dynamics of both the bulk and the groove water molecules. The important new result in this part of the work is the distribution of various relaxation times, such as of rotational correlation time. The distributions exhibit a log-normal nature signaling a unique character of groove water molecules as compared to the bulk.

In future, we aim to carry out longer simulations to understand the precise nature of the motion of the counterions. It seems that while the basic ideas of the continuous time random walk could remain valid, the incidence of queuing due to self-avoiding interaction might get weakened because of the availability of the extra space in bulk water. These aspects are under study.

## ACKNOWLEDGEMENTS

The authors thank the Department of Science and Technology (DST, India) for partial support of this work. B. Bagchi thanks Sir J. C. Bose fellowship for partial support. S. Mukherjee thanks DST, India for providing INSPIRE fellowship. S. Mondal thanks the University Grants Commission (UGC), India and S. Acharya thanks the Indian Institute of Science (IISc) for providing research fellowship.